\documentclass[11pt]{article}
\usepackage{easybmat}
\usepackage{cite}
\usepackage{amsmath}
\usepackage{apalike}
\usepackage{caption}
\usepackage{subcaption}
\usepackage{graphicx}
\usepackage{epsf,epsfig}
\usepackage[shortlabels]{enumitem}

\usepackage{savesym}
\savesymbol{AND}
\savesymbol{OR}
\savesymbol{NOT}
\savesymbol{TO}
\savesymbol{COMMENT}
\savesymbol{BODY}
\savesymbol{IF}
\savesymbol{ELSE}
\savesymbol{ELSIF}
\savesymbol{FOR}
\savesymbol{WHILE}
\usepackage{algorithmic}
\usepackage{algorithm}

\newtheorem{lemma}{Lemma}
\newtheorem{corollary}{Corollary}
\newtheorem{theorem}{Theorem}
\newtheorem{remark}{Remark}

\title{\bf Distributed Time Synchronization for Networks with Random Delays and Measurement Noise}

\author{Milo\v{s} S. Stankovi\'{c}\thanks{Innovation Center, School of Electrical Engineering, University of Belgrade,
Serbia, School of Technical Sciences, Singidunum University, Belgrade, Serbia, and the Vlatacom Institute, Belgrade, Serbia (e-mail: milstank@gmail.com).}, Srdjan S. Stankovi\'{c}\thanks{School of Electrical Engineering, University of Belgrade, Serbia and the Vlatacom Institute, Belgrade, Serbia (e-mail: stankovic@etf.rs).} and Karl Henrik Johansson \thanks{ACCESS Linnaeus Center, School of Electrical Engineering, KTH Royal Institute of Technology, Stockholm, Sweden (e-mail: kallej@kth.se).}  }

\begin{document}

\maketitle
\let\thefootnote\relax\footnote{This work was supported by the EU Marie Curie CIG (PCIG12-GA-2012-334098), Knut and Alice Wallenberg Foundation, the Swedish Strategic Research Foundation and the Swedish Research Council.}

\begin{abstract}
In this paper a new distributed asynchronous algorithm is proposed for time synchronization in networks with random communication delays, measurement noise and communication dropouts. Three different types of the drift correction algorithm are introduced, based on different kinds of local time increments. Under nonrestrictive conditions concerning network properties, it is proved that all the algorithm types provide convergence in the mean square sense and with probability one (w.p.1) of the corrected drifts of all the nodes to the same value (consensus). An estimate of the convergence rate of these algorithms is derived.  For offset correction, a new algorithm is proposed containing a compensation parameter coping with the influence of random delays and special terms taking care of the influence of both linearly increasing time and drift correction. It is proved that the corrected offsets of all the nodes converge in the mean square sense and w.p.1. An efficient offset correction algorithm based on consensus on local compensation parameters is also proposed. It is shown that the overall time synchronization algorithm can also be implemented as a flooding algorithm with one reference node. It is proved that it is possible to achieve bounded error between local corrected clocks in the mean square sense and w.p.1. Simulation results provide an additional practical insight into the algorithm properties and show its advantage over the existing methods.
\end{abstract}


\section{Introduction}

Cyber-Physical Systems (CPS), Internet of Things (IoT) and Sensor Networks (SN) have emerged as research areas of paramount importance, with many conceptual and practical challenges and numerous applications \cite{kim2012cyber,holler2014machine,akyildiz2010wireless}. One of the basic requirements in networked systems is, in general \emph{time synchronization}, \emph{i.e.}, all the nodes have to to share a common notion of time.
The problem of time synchronization has attracted a lot of attention, but still represents a challenge due to multi-hop communications, stochastic delays, communication and measurement noise, unpredictable packet losses and high probability of node failures, \emph{e.g.}, \cite{subuks}. There are numerous approaches to time synchronization starting from different assumptions and using different methodologies, \emph{e.g.}, \cite{elgies,subuks,siye}.  An important class of time synchronization algorithms
 is based on full
distribution of functions \cite{ssbs,soboku}. Distributed schemes with the so-called \textit{gradient
property}  have been proposed \cite{sowa,fl}. A class of \emph{consensus based algorithms}, called CBTS (Consensus-Based Time Synchronization) algorithms, has attracted considerable attention, \emph{e.g.}, \cite{liru,xk,he1,he2,scfi,liba,tian2}. It has been treated in a unified way in a recent survey \cite{tian1}, providing figure of merit of the principal approaches. In \cite{cczs1,ycs} a control-based approach to distributed time synchronization has been adopted.
 Fundamental and yet unsolved problems in all time synchronization approaches are connected with \emph{communication delays and measurement noise}; see \cite{fgk} for basic issues, and \cite{csq,xk,clsz,ggl1} for different aspects of delay influence.
\par
In this paper we propose a new \emph{asynchronous} \emph{distributed} algorithm for time synchronization in lossy networks, characterized by \emph{random }\emph{communication delays}, \emph{measurement noise} and \textit{communication dropouts}.
The algorithm is composed of two distributed recursions of \emph{asynchronous stochastic approximation type} based on \emph{broadcast gossip} and derived from predefined local error functions. The recursions are aimed at achieving asymptotic consensus on the \emph{corrected drifts} and \emph{corrected offsets} and, consequently, at obtaining \emph{common virtual clock} for all the nodes in the network.

 The proposed recursion for \emph{drift synchronization} (presented in a preliminary form in \cite{stankovic2016distributed}) is based on noisy time increments defined in three characteristic forms. We prove \emph{convergence to consensus} of the corrected drifts in the mean square sense and with probability one (w.p.1), under nonrestrictive conditions. Furthermore, we provide an estimate of the corresponding \emph{asymptotic convergence rate} to consensus. It is shown that the proposed recursion with the increments of unbounded length and with random boundaries provides the best performance, important for convergence to a common global virtual clock. Compared to the existing analogous algorithms \cite{scfi,tian2}, the proposed scheme is structurally different and simpler (not involving mutual drift estimation, typical for the CBTS algorithms) and, in addition, provides the best performance. Notice that the algorithm proposed in \cite{scfi} cannot handle communication delays and measurement noise, while the papers \cite{tian2,tian3}, derived from a particular form of increments of unbounded length, treat random delays, but not the case of measurement noise and communication dropouts. Moreover, the algorithm proposed therein cannot provide convergence rate achievable by the proposed methodology. The approach in \cite{ggl1} does not ensure consensus of corrected drifts in spite of additional pairwise inter-node communications.

We also propose a novel recursion for \emph{offset synchronization}, which starts from a specially constructed error function, derived from the difference between local times. Two important modifications are introduced, aiming at: 1) eliminating the deteriorating effect of linearly increasing absolute time, and 2) coping with the influence of delays by introducing additional \emph{delay compensation parameters}. It is proved that the algorithm provides convergence in the mean square sense and w.p.1 to a set of finite random variables.  The algorithm for the offset correction proposed in \cite{scfi} cannot handle these problems, while the algorithm in \cite{tian2,tian3} assume perfect clock readings. The approach in \cite{ycs} does not provide a rigorous insight into overall network stability.
Attention is also paid to an improvement of the offset correction algorithm, based on the introduction of linear consensus iterations in the recursion for the delay compensating parameters, aiming at decreasing the dispersion of the offset convergence points of different nodes. We believe that this modification can be a simple and efficient tool in practice. Special cases related to delay and noise are discussed in order to clarify potentials of the proposed algorithms.

The resulting time synchronization algorithm based on the proposed drift and offset correction recursions ensures finite differences between local virtual clocks in the mean square sense and w.p.1. To the authors' knowledge, the proposed algorithm represents the first method with such a performance in the case of random delays, measurement noise and communication dropouts.

It is also demonstrated that the proposed algorithm can be implemented as a \emph{flooding algorithm}, with one predefined reference node.

Finally, some illustrative simulation results are presented, giving an additional insight into the theoretically discussed issues.

\section{Synchronization Algorithms}

\subsection{Time and Network Models} Assume a network consisting of $n$ nodes, formally represented by a directed graph $ \mathcal{G}=(\mathcal{N},\mathcal{E})$, where
$\mathcal{N}$ is the set of nodes and $\mathcal{E}$ the set of arcs defining the structure of inter-node communications.
Denote by ${\mathcal N}^{+} _{i}$  the
 out-neighborhood and by ${\mathcal N}_{i}^{-}$ the in-neighborhood of node $i$, $i=1, \ldots, n$.
Assume that each node has a local clock, whose output, defining \emph{local time}, is given for any \emph{absolute time} $t \in \mathcal{R}$ by
\begin{equation} \label{yi}
\tau_{i}(t)=\alpha_{i} t + \beta_{i}+ \xi_{i}(t),
\end{equation}
where  $\alpha_{i} \neq 0$  is the local \emph{drift} \emph{(gain)}, $\beta_{i}$ is  the local \emph{offset}, while $\xi_{i}(t)$ is \emph{measurement noise}, appearing due to equipment instabilities, round-off errors, thermal noise, etc. \cite{liba,liba1,scfi,tsync2012}.
Each node $i$ applies an affine transformation to  $\tau_{i}(t)$, producing the \emph{corrected local time}
\begin{equation} \label{calfun}
\bar{\tau}_{i}(t)=a_{i} \tau_{i}(t) +b_{i} =  g_{i}t+ f_{i}+a_{i} \xi_{i}(t),
\end{equation}
where $a_{i}$ and $b_{i}$ are local \emph{correction parameters}, $g_{i}=a_{i} \alpha_{i}$ is the \emph{corrected drift} and $f_{i}=a_{i} \beta_{i}+ b_{i}$ the \emph{corrected offset}, $i=1, \ldots, n$.

The goal of distributed time synchronization is to provide a \emph{common virtual clock}, \emph{i.e.}, \emph{equal corrected drifts} $g_{i}$ and \emph{equal corrected offsets} $f_{i}$, $i=1, \ldots, n$, by \emph{distributed real-time estimation} of the parameters $a_{i}$ and $b_{i}$.
We assume that the nodes communicate according to the \emph{broadcast gossip scheme}, \emph{e.g.}, \cite{ned,aysal,bclz}, without global supervision or fusion center.
Namely, we assume that each node $j \in \mathcal{N}$ has its own \emph{local communication clock} that ticks  according to a Poisson process with the rate $\mu_{j}$, independently of other nodes. At each tick of its communication clock (denoted by $t_b^j$, $b=0,1,2, \ldots$), node $j$ broadcasts its current local time (together with its current estimates of the correction parameters) to its out-neighbors $i \in \mathcal{N}_{j}^{+}$. Each node $i \in \mathcal{N}_{j}^{+}$ hears the broadcast with probability $p_{ij} > 0$. Let $\{t_{l}^{j,i}\}$, $l=0,1,2, \ldots$, be the sequence of absolute time instants corresponding to the messages heard by node $i$. The message sent at $t_{l}^{j,i}$ is received at node $i$ at the time instant \[\bar{t}_{l}^{j,i}=t^{j,i}_{l}+ \delta^{j,i}_{l},\]
where $\delta^{j,i}_{l}$ represents the corresponding \emph{communication delay}. See \cite{xk,lewu,csq,fgk,clsz} for presentation of physical and technical sources of the delays. We assume in the sequel that the communication delay can be decomposed as
\begin{equation} \label{delta}
\delta_{l}^{j,i}= \bar{\delta}^{j,i}+ \eta_{i}(\bar{t}_{l}^{j,i}),
\end{equation}
where $\bar{\delta}^{j,i}$ is assumed to be constant (depending only on the chosen arc $(j,i)$), while $\eta_{i}(\bar{t}_{l}^{j,i})$ represents a stochastically time-varying component with zero mean. After receiving a message from node $j$, node $i$ reads its current local time, calculates its own current \emph{corrected local time} and \emph{updates the values of its correction parameters} $a_{i}$ and $b_{i}$. The process is repeated after each tick of the communication clock of any node in the network; we assume, as usually, that time is dense and only one communication clock can tick at a given time \cite{ned}.

\subsection{Drift Correction Algorithm} The drift correction algorithm is given as Algorithm~\ref{alg1} and described in the following. The recursion for updating the value of parameter $a_{i}$ at node $i$, as a response to a message coming from node $j$, is based on the following \emph{error function}:
\begin{equation} \label{ja}
\bar{\varphi}^{a}_{i} (\bar{t}_{l}^{j,i})=  \Delta \bar{\tau} _{j}(t_{l}^{j,i})-
\Delta \bar{\tau} _{i}(\bar{t}_{l}^{j,i}),
\end{equation}
where $\Delta \bar{\tau} _{j}(t_{l}^{j,i})$ and
$\Delta \bar{\tau} _{i}(\bar{t}_{l}^{j,i})$ are \emph{increments of the corrected local times}, given by
 \[\Delta \bar{\tau} _{j}(t_{l}^{j,i})=\bar{\tau} _{j}(t_{l}^{j,i})- \bar{\tau} _{j}(t_{m}^{j,i})=a_{j}  \Delta \tau_{j}(t_{l}^{j,i}), \] \[\Delta \bar{\tau} _{i}(\bar{t}_{l}^{j,i})=\bar{\tau} _{i}(\bar{t}_{l}^{j,i})- \bar{\tau} _{i}(\bar{t}_{m}^{j,i})=a_{i} \Delta \tau_{i} (\bar{t}_{l}^{j,i}), \] where $m \in \{0, \ldots, l-1 \}$,
 \[\Delta \tau_{j}(t_{l}^{j,i})= \tau_{j}(t_{l}^{j,i})- \tau_{j}(t_{m}^{j,i})= \alpha_{j} \Delta t_{l}^{j,i}+ \Delta \xi_{j}(t_{l}^{j,i}), \]
 \[ \Delta \tau_{i}(\bar{t}_{l}^{j,i})= \alpha_{i} \Delta \bar{t}_{l}^{j,i}+ \Delta \xi_{i}(\bar{t}_{l}^{j,i}),\]
$\Delta t_{l}^{j,i}=t_{l}^{j,i}-t_{m}^{j,i}$,  $ \Delta \xi_{j}(t_{l}^{j,i})=\xi_{j}(t_{l}^{j,i})-\xi_{j}(t_{m}^{j,i})$, $\Delta \bar{t}_{l}^{j,i}=\bar{t}_{l}^{j,i}-\bar{t}_{m}^{j,i}= \Delta t_{l}^{j,i} +  \Delta \delta_{l}^{j,i},$ with $\Delta \delta_{l}^{j,i}= \delta_{l}^{j,i}-  \delta_{m}^{j,i}$, and $\Delta \xi_{i}(\bar{t}_{l}^{j,i})= \xi_{i}(\bar{t}_{l}^{j,i})-\xi_{i}(\bar{t}_{m}^{j,i})$; by (\ref{delta}), we have $\Delta \delta_{l}^{j,i}=\Delta \eta_{i}(\bar{t}_{l}^{j,i})$, where $\Delta \eta_{i}(\bar{t}_{l}^{j,i})= \eta_{i}(\bar{t}_{l}^{j,i})-\eta_{i}(\bar{t}_{m}^{j,i})$.

Here $m$ denotes the index of the past time instant with respect to which the time increment is calculated. The choice of $m$ leads to different definitions of the time increment, and to algorithms with different properties. In this paper we shall consider the following three characteristic cases (which we denote as \emph{AlgDrift.a}, \emph{AlgDrift.b} and \emph{AlgDrift.c}; see Algorithm~\ref{alg1}):

\begin{enumerate}[a)]
\item $m=l-L$, where $L>0$ is a predefined integer (\emph{AlgDrift.a});
\item $m=\lfloor\nu l \rfloor$ ($0 < \nu < 1$), where $\lfloor x\rfloor$ denotes the largest integer less than or equal to $x$ (\emph{AlgDrift.b});
\item $m=l_{0}$, where $l_{0}$ is a fixed integer (\emph{AlgDrift.c}).
\end{enumerate}

\begin{remark}
In \emph{AlgDrift.a} and \emph{AlgDrift.c} the required memory is finite; in \emph{AlgDrift.a} the memory requirement is determined by $L$ (in the algorithm proposed in \cite{scfi} $L=1$). In \emph{AlgDrift.b} and \emph{AlgDrift.c} the increment length is unbounded. \emph{AlgDrift.c} is based on the idea first formulated in \cite{tian1,tian2,tian3} using a fixed initial time instant $m=l_{0}$. However, in \emph{AlgDrift.b} we have both $\lim_{l \to \infty} m =\infty$ and $\lim_{l \to \infty}(l- m) =\infty$, which is conceptually essentially important since it provides the highest convergence rate (see Theorem~\ref{th2} below), and indicates that the best scheme for practice is obtainable by choosing \emph{AlgDrift.a} with $L$ large enough (see simulations Section~\ref{sim}).
\end{remark}

Using (\ref{ja}) we define the following updating procedure for parameter $a_{i}$ at node $i$,
to be executed immediately after node $i$  receives the message from node $j$ ($j=1, \ldots, n$, $i \in {\mathcal N}^{+} _{j}$):
\begin{equation} \label{ai}
\hat{a}_{i}(\bar{t}_{l}^{j,i \, +})=\hat{a}_{i}(\bar{t}_{l}^{j,i})+ \varepsilon_{i}^a(\bar{t}_{l}^{j,i}) \gamma_{ij} \hat{\varphi}^{a}_{i}(\bar{t}_{l}^{j,i}),
\end{equation}
where $\gamma_{ij}$ are \emph{a priori} chosen nonnegative weights expressing relative importance of communication links (their role will be discussed below),
$\hat{\varphi}^{a}_{i} (\bar{t}_{l}^{j,i})=\Delta \hat{\tau}_{j}(t_{l}^{j,i})-\Delta \hat{\tau}_{i}(\bar{t}_{l}^{j,i}),$
\begin{align}
\Delta \hat{\tau}_{j}(t_{l}^{j,i})&= \Delta \bar{\tau}_{j}(t_{l}^{j,i})|_{a_{j}=\hat{a}_{j}(t_{l}^{j,i})}, \label{delta_hat_tau_j} \\
\Delta \hat{\tau}_{i}(\bar{t}_{l}^{j,i})&= \Delta \bar{\tau}_{i}(\bar{t}_{l}^{j,i})|_{a_{i}=\hat{a}_{i}(\bar{t}_{l}^{j,i})}, \label{delta_hat_tau_i}
\end{align}
$\hat{a}_{j}(t_{l}^{j,i})$ and $ \hat{a}_{i}(\bar{t}_{l}^{j,i})$ are the old estimates, $\hat{a}_{i}(\bar{t}_{l}^{j,i \, +})$ the new estimate, while
$\varepsilon_{i}^a(\bar{t}_{l}^{j,i})$ is a positive step size.  The corresponding pseudocode is presented as Algorithm~\ref{alg1}. The updating procedure (\ref{ai}) generates, in such a way, recursions of \emph{distributed asynchronous stochastic approximation} type. It will be assumed that the initial estimates are  $\hat{a}_{i}(\bar{t}_{0}^{j,i})=1$.

\begin{algorithm}
\caption{\textit{AlgDrift.a}, \textit{AlgDrift.b} and \textit{AlgDrift.c}}
\label{alg1}
\begin{algorithmic}
\FOR {All the nodes $i\in\mathcal{N}$}
\STATE{Initialize $\hat{a}_{i}(\bar{t}_{0}^{j,i})=1$}
\ENDFOR
\LOOP
\IF {Tick $t_b^j$ of a local communication clock of a node $j\in\mathcal{N}$}

\STATE {Read the current local time value $\tau_j(t_b^j)$}
\STATE {Broadcast $\tau_j(t_b^j)$ and $\hat{a}_{j}(t_b^j)$ to the out-neighbours $\mathcal{N}_{j}^{+}$}

\ENDIF
\ENDLOOP

\LOOP
\IF {A message received by a node $i\in\mathcal{N}$ from a node $j\in\mathcal{N}_{j}^{-}$ (at absolute time $\bar{t}_l^{j,i}$)}

\IF {The first message from the node $j$}
\STATE {Save the received initial local time of node $j$ $\tau_j(t_0^{j,i})$}
\STATE {Read and save the initial local time $\tau_i(\bar{t}_0^{j,i})$}
\ELSE
\STATE {Read the current local time value $\tau_i(\bar{t}_l^{j,i})$}
\STATE {Calculate $\Delta \hat{\tau}_j(t_l^{j,i})$ and $\Delta \hat{\tau}_i(\bar{t}_l^{j,i})$ according to \eqref{delta_hat_tau_j} and \eqref{delta_hat_tau_i}, where $m=l-L$ for \textit{AlgDrift.a}, $m=\lfloor\nu l \rfloor$ for \textit{AlgDrift.b}, and $m=0$ for \textit{AlgDrift.c}}
\STATE {Calculate a new estimate of the drift correction parameter according to \eqref{ai}}
\ENDIF

\ENDIF
\ENDLOOP
\end{algorithmic}
\end{algorithm}

In terms of the corrected drift $\hat{g}_{i}(\cdot)= \hat{a}_{i}( \cdot) \alpha_{i}$, (\ref{ai}) gives:
\begin{equation} \label{gi}
\hat{g}_{i}(\bar{t}_{l}^{j,i \, +})=\hat{g}_{i}(\bar{t}_{l}^{j,i})+ \varepsilon_{i}^a(\bar{t}_{l}^{j,i}) \gamma_{ij} \hat{\psi}^{a}_{i} (\bar{t}_{l}^{j,i}),
\end{equation}
where
\begin{align} 
\hat{\psi}^{a}_{i} (\bar{t}_{l}^{j,i})=&\alpha_{i}
\{[\hat{g}_{j}(t_{l}^{j,i})-\hat{g}_{i}(\bar{t}_{l}^{j,i})] \Delta t^{j,i}_{l}
 +\frac{1}{\alpha_{j}}\hat{g}_{j} (t_{l}^{j,i})\Delta \xi_{j}(t_{l}^{j,i}) \nonumber \\ &-  \frac{1}{\alpha_{i}} \hat{g}_{i} (\bar{t}_{l}^{j,i}) \Delta \xi_{i}(\bar{t}_{l}^{j,i})
-\hat{g}_{i}(\bar{t}_{l}^{j,i}) \Delta \eta_{i}(\bar{t}_{l}^{j,i}) \}.
\end{align}

\begin{remark} The basic drift correction estimation scheme \eqref{ai},\eqref{gi} is independent of offset correction, with a role analogous to the distributed drift estimation schemes in \cite{scfi,tian1,tian2}. However, it does not belong to the class of the so called CBTS algorithms \cite{tian1}: it is structurally different and simpler, not requiring the step of \emph{relative drift} estimation, which introduces unnecessary dynamics and additional nonlinearities. Even a robustified version of the algorithm in \cite{scfi} proposed in \cite{ggl1} cannot achieve consensus of corrected drifts in the case of stochastic delays.
\end{remark}
\begin{remark}
For $L=1$ in \emph{AlgDrift.a} and $l_{0}=0$ in \emph{AlgDrift.c} one obtains the drift correction algorithms proposed in \cite{stankovic2016distributed}. Within the context of CBTS algorithms, $m=l-1$ has been used in \cite{scfi}, and $m=l_{0}$ in \cite{tian1,tian2,tian3}. A pseudo periodic version of (\ref{ai}) with $m=l-1$ has been proposed and analyzed in \cite{tsync2012}.
\end{remark}

\subsection{Offset Correction Algorithm} The offset correction algorithm is given as Algorithm~\ref{alg2} and described below. The recursion for updating parameter $b_{i}$ is based on the following error function:
\begin{align} \label{jb}
 \bar{\varphi}_{i}^{b} (\bar{t}_{l}^{j,i})= & \bar{\tau} _{j}(t_{l}^{j,i})- a_{j} T_{j}(t_{l}^{j,i})-(\bar{\tau} _{i}(\bar{t}_{l}^{j,i}) - a_{i} T_{i}(\bar{t}_{l}^{j,i})) + c_{i},
\end{align}
$j=1, \ldots, n$, $i \in \mathcal{N}_{j}^{+}$,
where
\begin{equation} \label{Tij}
T_{j}(t_{l}^{j,i})=\Delta \tau_{j}(t_{l}^{j,i})|_{m=0}, \; T_{i}(\bar{t}_{l}^{j,i})=\Delta \tau_{i}(\bar{t}_{l}^{j,i})|_{m=0},
\end{equation}
while $c_{i}$ is an additional \emph{delay compensation parameter}. \par
Using (\ref{jb}), we come up with the following updates for $b_{i}$ and $c_{i}$ (based on the assumption that an estimate of $a_{i}$ is given):
\begin{equation}  \label{bi}
\hat{b}_{i}(\bar{t}_{l}^{j,i \, +})=\hat{b}_{i}(\bar{t}_{l}^{j,i})+ \varepsilon_i^b(\bar{t}_{l}^{j,i}) \gamma_{ij} \hat{\varphi}_{i}^{b} (\bar{t}_{l}^{j,i})
\end{equation}
\begin{equation}  \label{ci}
\hat{c}_{i}(\bar{t}_{l}^{j,i \, +})=\hat{c}_{i}(\bar{t}_{l}^{j,i}) - \varepsilon_i^b(\bar{t}_{l}^{j,i}) \gamma_{ij} \hat{\varphi}_{i}^{b} (\bar{t}_{l}^{j,i})
\end{equation}
where
$
\hat{\varphi}_{i}^{b} (\bar{t}_{l}^{j,i})= \hat{\tau}_{j}(t_{l}^{j,i})- \hat{a}_{j}(t_{l}^{j,i}) T_{j}(t_{l}^{j,i}) $ $-(\hat{\tau}_{i}(\bar{t}_{l}^{j,i}) -\hat{a}_{i}(\bar{t}_{l}^{j,i}) T_{i}(\bar{t}_{l}^{j,i}))+ \hat{c}_{i}(\bar{t}_{l}^{j,i})$, with
\begin{align}
\hat{\tau}_{j}(t_{l}^{j,i})&=\hat{a}_{j}(t_{l}^{j,i})\tau_{j}(t_{l}^{j,i}) + \hat{b}_{j}(t_{l}^{j,i}), \label{hat_tau_j} \\
\hat{\tau}_{i}(\bar{t}_{l}^{j,i})&=\hat{a}_{i}(\bar{t}_{l}^{j,i})\tau_{i}(\bar{t}_{l}^{j,i}) + \hat{b}_{i}(\bar{t}_{l}^{j,i}). \label{hat_tau_i}
\end{align}
The initial estimates are supposed to be $\hat{b}_{i}(\bar{t}_{0}^{j,i})=0$ and $\hat{c}_{i}(\bar{t}_{0}^{j,i})=0$. The estimates of the drift correction parameters can be generated by any convenient algorithm; when it is generated by (\ref{ai}), we obtain a complete new time synchronization algorithm.

In terms of $\hat{g}_{i}(\cdot)= \hat{a}_{i}( \cdot) \alpha_{i}$ and
$\hat{f}_{i}(\cdot) = \hat{a}_{i}( \cdot) \beta_{i}+ \hat{b}_{i}( \cdot)$, (\ref{bi}) and (\ref{ci}) become:
\begin{equation}  \label{fi}
\hat{f}_{i}(\bar{t}_{l}^{j,i \, +})+ \Delta \hat{g}_{i}(\bar{t}_{l}^{j,i \, +})=\hat{f}_{i}(\bar{t}_{l}^{j,i})+ \varepsilon_{i}^b(\bar{t}_{l}^{j,i}) \gamma_{ij} \hat{\psi}^{b}_{i} (\bar{t}_{l}^{j,i}),
\end{equation}
\begin{equation}  \label{cii}
\hat{c}_{i}(\bar{t}_{l}^{j,i \, +})=\hat{c}_{i}(\bar{t}_{l}^{j,i})- \varepsilon_{i}^b(\bar{t}_{l}^{j,i}) \gamma_{ij} \hat{\psi}^{b}_{i} (\bar{t}_{l}^{j,i}),
\end{equation}
where $ \Delta \hat{g}_{i}(\bar{t}_{l}^{j,i \, +})= \frac{\beta_{i}}{\alpha_{i}}[\hat{g}_{i}(\bar{t}_{l}^{j,i})-\hat{g}_{i}(\bar{t}_{l}^{j,i \, +})]$ and
\begin{align}
\hat{\psi}^{b}_{i} (\bar{t}_{l}^{j,i})=&
[\hat{g}_{j}(t_{l}^{j,i})-\hat{g}_{i}(\bar{t}_{l}^{j,i})] t_{0}^{j,i} + \hat{f}_{j} (t_{l}^{j,i}) - \hat{f}_{i} (\bar{t}_{l}^{j,i})-\hat{g}_{i} (\bar{t}_{l}^{j,i}) [\bar{\delta}^{i,j}+\eta_{i}(\bar{t}_{0}^{j,i})] \nonumber \\&+\hat{c}_{i}(\bar{t}_{l}^{j,i})+
\frac{1}{\alpha_{j}} \hat{g}_{j} (t_{l}^{j,i}) \xi_{j}(t_{0}^{j,i})-\frac{1}{\alpha_{i}} \hat{g}_{i} (\bar{t}_{l}^{j,i}) \xi_{i}(\bar{t}_{0}^{j,i}).
\end{align}

A \emph{consensus-based modification} of (\ref{ci}) and (\ref{cii}) will be considered apart. This modification is formally obtained by replacing $\hat{c}_{i}(\bar{t}_{l}^{j,i})$ at the right hand side of (\ref{ci}) and (\ref{cii}) by the following \emph{convex combination}
\begin{equation} \label{consensus}
\hat{c}^{con}_{i}(\bar{t}_{l}^{j,i})= \sigma_{i}\hat{c}_{i}(\bar{t}_{l}^{j,i})+ (1-\sigma_{i}) \hat{c}_{j}(\bar{t}_{l}^{j,i}),
\end{equation}
with tuning parameter $0 < \sigma_{i} \leq 1$. This modification is motivated by a realistic assumption that the delays in the network are not too far from each other, with the aim to achieve smaller dissipation of the convergence points for $\hat{f}_{i}(\cdot) $ (see Remark~\ref{cons} below). We refer to this algorithm as \textit{AlgOffset.b}. The pseudocode of the algorithms \textit{AlgOffset.a} and \textit{AlgOffset.b} are presented as Algorithm~\ref{alg2}.

\begin{algorithm}
\caption{\textit{AlgOffset.a} and \textit{AlgOffset.b}}
\label{alg2}
\begin{algorithmic}
\FOR {All the nodes $i\in\mathcal{N}$}
\STATE{Initialize $\hat{b}_{i}(\bar{t}_{0}^{j,i})=0$ and $\hat{c}_{i}(\bar{t}_{0}^{j,i})=0$}
\ENDFOR

\LOOP
\IF {Tick $t_b^j$ of a local communication clock of a node $j\in\mathcal{N}$}

\STATE {Read the current local time value $\tau_j(t_b^j)$}
\STATE {Broadcast $\tau_j(t_b^j)$, $\hat{a}_{j}(t_b^j)$, $\hat{b}_{j}(t_b^j)$ and $\hat{c}_{j}(t_b^j)$ to the out-neighbours $\mathcal{N}_{j}^{+}$}

\ENDIF
\ENDLOOP

\LOOP
\IF {A message received by a node $i\in\mathcal{N}$ from a node $j\in\mathcal{N}_{j}^{-}$ (at absolute time $\bar{t}_l^{j,i}$)}

\STATE {Read the current local time value $\tau_i(\bar{t}_l^{j,i})$}
\STATE {Calculate $\hat{\tau}_{j}(t_{l}^{j,i})$ and $\hat{\tau}_{i}(\bar{t}_{l}^{j,i})$ using \eqref{hat_tau_j} and \eqref{hat_tau_i}} \STATE {Calculate $T_{j}(t_{l}^{j,i})$ and $T_{i}(\bar{t}_{l}^{j,i})$ using \eqref{Tij}}
\STATE {Calculate new estimates of the offset correction parameters according to \eqref{bi} and \eqref{ci} (and \eqref{consensus} for the algorithm \textit{AlgOffset.b})}

\ENDIF
\ENDLOOP
\end{algorithmic}
\end{algorithm}

\begin{remark} \label{rem_offset} The proposed offset estimation scheme represented by (\ref{bi})--(\ref{cii}) is based on \emph{two major modifications} of the basic error function $\varphi^{b}_{i} (\bar{t}_{l}^{j,i})^{0}= \bar{\tau}_{j}(t_{l}^{j,i}) -\bar{\tau}_{i}(\bar{t}_{l}^{j,i})$, which has been utilized in all the existing CBTS algorithms (see \cite{tian1} and the references therein). The \emph{first modification} introduces two easily computable terms $T_{j}(t_{l}^{j,i})$ and $T_{i}(\bar{t}_{l}^{j,i})$, the role of which is to cope with the unboundedly increasing term $t_{l}^{j,i}$ in the expression for $\varphi^{b}_{i} (\bar{t}_{l}^{j,i})^{0}$, in such a way as to replace it with the bounded term $t_{0}^{j,i}$ in $\varphi^{b}_{i} (\bar{t}_{l}^{j,i})$. The \emph{second modification} consists of introducing a \emph{new parameter} to be updated, $\hat{c}_{i}(\bar{t}_{l}^{j,i})$. Estimation of this parameter is aimed at coping directly with the effects of communication delays and enabling convergence of the offset correction parameter estimates.
\end{remark}
\begin{remark} The proposed time synchronization algorithm requires very small communication and computation efforts. At each tick $t_{l}^{j,i}$, a packet is sent by the $j$-th node to
its neighbors $i \in \mathcal{N}_{j}^{+}$, containing the current local time $\tau_{j}(t_{l}^{j,i})$ and the current local drift and offset correction parameter estimates $\hat{a}_{j}(t_{l}^{j,i})$, $\hat{b}_{j}(t_{l}^{j,i})$ and $\hat{c}_{j}(t_l^{j,i})$. After
receiving this packet, the neighbors calculate the corresponding $\Delta \hat{\tau}_{j}(t_{l}^{j,i})$, $\Delta \hat{\tau}_{i}(\bar{t}_{l}^{j,i})$, $ \hat{\tau}_{j}(t_{l}^{j,i})$, $ \hat{\tau}_{i}(\bar{t}_{l}^{j,i})$, $T_{j}(t_{l}^{j,i})$ and $T_{i}(\bar{t}_{l}^{j,i})$, and update their own parameter estimates according to \eqref{ai}, \eqref{bi}, \eqref{ci} and \eqref{consensus}. The same procedure is repeated after each new tick of any of the nodes.
\end{remark}

\subsection{Global Model}

Next, we derive a global model of the overall time synchronization network. Parameter updating at the network level is driven by a \emph{global virtual communication clock}, with the rate equal to $\mu_{c}=\sum_{i=1}^{n} \mu_{i}$, that ticks whenever any of the local communication clocks tick (\emph{e.g.}, \cite{ned,aysal}). Starting from this fact, a global model for the whole network has been defined in \cite{stankovic2016distributed} in the form of a recursion in which the iteration number corresponds to the number of a tick of the global virtual communication clock. In this paper, we shall adopt an alternative approach, providing a more direct insight into the whole parameter updating process. Namely, we shall assume that each local update in the network produces a unique iteration number $k$ in the global model of the parameter estimates, and, \textit{vice versa}, that each $k$ is connected to an update of $i$-th node (the corresponding continuous time instant is $\bar{t}^{j,i}_{l}$ for some  $j$ and $l$). In such a way, at a click of $j$-th communication clock we have $N(j)$ consecutive updates or iterations (assuming that we have only one update at a time), $N(j) \leq |\mathcal{N}^{+}_{j}|$. Following analogous approaches in \cite{ned,tian1}, we replace (with some abuse of notation) the variable $\bar{t}^{j,i}_{l}$ by $k$ in all the above defined functions of time, so that we have $\tau_{i}(\bar{t}^{j,i}_{l})=\tau_{i}(k)$, $\bar{\tau}_{i}(\bar{t}^{j,i}_{l})=\bar{\tau}_{i}(k)$, $\xi_{i}(\bar{t}_{l}^{j,i})= \xi_{i}(k)$, etc; accordingly, we also write $\tau_{j}(t^{j,i}_{l})=\tau_{j}(k)$, $\bar{\tau}_{j}(t^{j,i}_{l})=\bar{\tau}_{j}(k)$, $\xi_{j}(t_{l}^{j,i})= \xi_{j}(k)$, etc. In the case of delays, we write $\bar{\delta}^{j,i}=\bar{\delta}_{j}(k)$ and $\eta_{i}(\bar{t}^{j,i}_{l})$ $=\eta_{i}(k)$.

Assume that $k$ is connected to an update at node $i$, initiated by a tick of node $j$. Let
 $\hat{g}(k)=
[\hat{g}_{1}(k) \ldots \hat{g}_{n}(k)]^{T}$,  $\hat{f}(k)=
[\hat{f}_{1}(k) \ldots \hat{f}_{n}(k)]^{T}$ and $\hat{c}(k)=[\hat{c}_{1}(k) \ldots \hat{c}_{n}(k)]^{T}$,
where $\hat{g}_{\mu}(k)=\hat{a}_{\mu}(k) \alpha_{\mu}$, $ \hat{a}_{\mu}(k)=\hat{a}_{\mu}(\bar{t}_{l}^{j,i})$,  $\hat{f}_{\mu}(k)=\hat{a}_{\mu}(k) \beta_{\mu}+ \hat{b}_{\mu}(k)$, $\hat{b}_{\mu}(k)=\hat{b}_{\mu}(\bar{t}_{l}^{j,i})$ and $\hat{c}_{\mu}(k)=\hat{c}_{\mu}(\bar{t}_{l}^{j,i})$,  $\mu=1, \ldots, n$,
 Then, (\ref{gi}) gives
\begin{equation} \label{algg}
\hat{g}(k+1)= \hat{g}(k)   +\varepsilon^a(k)  Z(k) \hat{g}(k),
\end{equation}
where $\hat{g}(k+1) =[\hat{g}_1(\bar{t}^{j,1+}_l) \ldots \hat{g}_n(\bar{t}^{j,n+}_l)]^T$,
$ \varepsilon^a(k)={\rm diag} \{\varepsilon_{1}^a(k), \ldots, \varepsilon_{n}^a(k) \}$, $\varepsilon_{i}^a(k)=\varepsilon_{i}^a(\bar{t}^{j,i}_{l})$ (see (\ref{ai})), \[Z(k)=A \Gamma(k) \Delta t(k)+ N_{g} (k),\]
$A= {\rm diag} \{ \alpha_{1}, \ldots, \alpha_{n} \},$
$ \Gamma(k)= [\Gamma(k)_{\mu \nu}],$ with $\Gamma(k)_{ i i}= -\gamma_{i j}$ and $\Gamma(k)_{i j}= \gamma_{i j}$, with $\Gamma(k)_{\mu \nu}=0$ otherwise, $\Delta t(k)=\bar{t}_{l}^{j,i}-\bar{t}_{m}^{j,i}, $  while the noise term is defined as
\[N_{g}(k)= - A \Gamma_{d}(k) \Delta \eta_{d} (k)
+  A \Gamma(k) \Delta \xi_{d}(k) A^{-1}, \]
 where $\Gamma_{d}(k)= {\rm diag} \{{\rm diag} \{ \gamma_{1j}, \ldots, \gamma_{nj}\} \omega(k)\} $, \newline  $ \omega(k)=[\omega_{1}(k) \ldots \omega_{n}(k)]^{T}$, $\omega_{i}(k)=1$, $\omega_{\mu }(k)=0$ for $\mu \neq i$,  $\Delta \eta_{d}(k)={\rm diag} \, \Delta \eta(k)$,
$\Delta \eta(k)=[\Delta \eta_{1}(k) \ldots \Delta \eta_{n}(k)]^{T}$,  $\Delta \xi_{d} (k)={\rm diag} \, \Delta \xi(k)$ and $ \Delta \xi(k)=[\Delta \xi_{1}(k) \ldots \Delta \xi_{n}(k)]^{T}$.
\par
Similarly, from
(\ref{fi}) and (\ref{cii}) we obtain
\begin{align}
\hat{f}(k+1)+ \Delta \hat{g}(k+1)&=\hat{f}(k)+\varepsilon^b(k) Y(k) \label{algf} \\
\hat{c}(k+1)&=\hat{c}(k)-\varepsilon^b(k) Y(k),  \label{algc}
\end{align}
where $\Delta \hat{g}(k+1)={\rm diag} \, \omega(k)(\hat{g}(k+1)-\hat{g}(k))$,  $ Y(k)=\Gamma(k) \hat{f}(k) + [t^{0}(k) \Gamma(k)- $ $\Gamma_{d}(k)  \bar{\delta}_{d}(k) -\Gamma_{d}(k)  \eta_{d}^{0}(k) $ $ +\Gamma(k) \xi_{d}^{0}(k) A^{-1}] \hat{g}(k)
+ \Gamma_{d}(k)  \hat{c}(k),$
$t^{0}(k)=t_{0}^{j,i}$,  $\bar{\delta}_{d}(k)= {\rm diag} \, \bar{\delta}(k),\; \bar{\delta}(k)=[\bar{\delta}^{j,1} \ldots \bar{\delta}^{j,n}]^{T}$, $\eta_{d}^{0}(k)={\rm diag} \, \eta^{0}(k), \;$ $ \eta^{0}(k)=[\eta_{1}^{0}(k) \ldots \eta_{n}^{0}(k)]^{T}, $ ($\eta_{i}^{0}(k) = \eta_{i}(\bar{t}^{j,i}_{0})$),  $\xi_{d}^{0}(k)={\rm diag} \, \xi^{0}(k),
\xi^{0}(k)= [\xi_{1}^{0}(k) \ldots \xi_{n}^{0}(k)]^{T}$ ($\xi_{j}^{0}(k)=\xi_{j}(t^{j,i}_{0})$, $\xi_{i}^{0}(k)=\xi_{i}(\bar{t}^{j,i}_{0})$; $\{t^{0}(k) \}$, $\{ \eta^{0}(k)\}$ and $\{ \xi^{0}(k)\}$ are random sequences with finite sets of possible realizations composed of $t^{j,i}_{0}$, $\eta_{i}(\bar{t}^{j,i}_{0})$ and $\xi_{j}(t^{j,i}_{0})$ (or $\xi_{i}(\bar{t}^{j,i}_{0})$), obtained at each $k$ by choosing $j$ and $i$ at random.
\par
In \textit{AlgOffset.b}, $\hat{c}(k)$ is replaced by $\hat{c}^{con}(k)=C(k) \hat{c}(k) $, where $C(k)= [C(k)_{\mu \nu}],$  with $C(k)_{ \mu \mu}= \sigma_{\mu}$ and $C(k)_{\mu j}= 1-\sigma_{\mu}$ for all $\mu \in {\mathcal N}^{+} _{j}$, with $C(k)_{\mu \nu}=0$ otherwise.

\section{Convergence Analysis} \label{sec3}

\subsection{Preliminaries}
\par
Within the exposed general setting, we additionally assume:
\par
(A1) Graph $\mathcal{G}$ has a spanning tree.
\par
(A2)  $ \{\xi_{i}(k) \}$ and $ \{\eta_{i}(k) \}$ , $i=1, \ldots n$, are mutually independent zero mean i.i.d. random sequences, bounded w.p.1.
\par
(A3)  The step sizes $\varepsilon_{i}^a(k)$ and $\varepsilon_{i}^b(k)$ are defined in the following way: \newline $\varepsilon_{i}^a(k) =\varepsilon_{i}(k)|_{\zeta=\zeta'}$ for \emph{AlgDrift.a}, \newline  $\varepsilon_{i}^a(k) =\varepsilon_{i}(k)|_{\zeta=1+\zeta'}$ for \emph{AlgDrift.b} and \emph{AlgDrift.c},  and \newline $\varepsilon_{i}^b(k) =\varepsilon_{i}(k)|_{\zeta=\zeta''}$ for \emph{AlgOffset.a} and \emph{AlgOffset.b}, \newline where
 $\varepsilon_{i}(k)= \nu_{i}(k)^{-\zeta}$, $\nu_{i}(k)= \sum_{m=1}^{k} I\{$ node $i$ received a message$\}$, representing the number of updates of node $i$ up to the instant $k$ ($I\{\cdot\}$ denotes the indicator function), while  $\frac{1}{2} < \zeta',\zeta'' \leq 1$.

\begin{remark} (A1) implies that graph $\mathcal{G}$ has a center node from which all the remaining nodes are reachable \cite{osfm,calieee}. (A2) is a standard assumption, in which boundedness, which always holds in practice, is introduced for making derivations easier. (A3) is practically very important: it eliminates the need for a centralized clock which would define the common step size for all the nodes as a function of $k$. The choice of the exponent in the expression for $\varepsilon_{i}^a(k)$ for \emph{AlgDrift.b} and \emph{AlgDrift.c} is motivated by the properties of the corresponding random variable  $\Delta t(k)$ which diverges linearly to infinity (see Theorem~\ref{th2}). The choice of $\zeta'$ and $\zeta''$ is standard for stochastic approximation algorithms.
\end{remark}

Asymptotical behavior of the step size is given by the following lemma. Proofs of all the lemmas and theorems are given in the Appendix.
\par
\begin{lemma} \label{lm1} Let (A1) and (A3) be satisfied, let $p_{i}$ be the unconditional probability of node $i$ \emph{to update} its parameters at $k$-th iteration, and let $\zeta>0$. Then, for a given $q' \in (0, \frac{1}{2})$, there exists an integer $\bar{k}>0$ such that w.p.1 for all $k \geq \bar{k}$
\begin{equation} \label{Delta}
\varepsilon_{i}(k)=\frac{1}{k^{\zeta}} \left(\frac{\bar{N}}{p_{i}}\right)^{\zeta} + \tilde{\varepsilon}_{i}(k),
\end{equation}
where $\bar{N} =E_{j} \{ E \{ N(j)|j \} \}$ represents the average number of updates per one tick of the global virtual clock, and
$
|\tilde{\varepsilon}_{i}(k)| \leq \tilde{\varepsilon}_{i} \frac{1}{k^{\zeta+\frac{1}{2}-q'}},
$
 $0 < \tilde{\varepsilon}_{i} < \infty$, $i=1, \ldots, n$.
\end{lemma}

Properties of the matrix $\Gamma(k)$ defined in the previous section are essential for convergence of (\ref{algg})--(\ref{algc}); its expectation $\bar{\Gamma}= E\{ \Gamma (k) \}$ has the central role in the analysis, since it contains all the information about the network structure and the weights of particular links. It has the structure of a weighted Laplacian matrix for $\mathcal{G}$:
\begin{equation} \label{Laplacian}
 \bar{\Gamma} =\left[
\begin{BMAT}{cccc}{cccc} -\sum_{j, j \neq 1} \gamma_{1j} \pi_{1j}  &  \gamma_{12} \pi_{12} & \cdots & \gamma_{1n} \pi_{1n} \\ \gamma_{21} \pi_{21} & -\sum_{j, j \neq 2}
\gamma_{2j} \pi_{2j}& \cdots & \gamma_{2n} \pi_{2n} \\ & & \ddots & \\ \gamma_{n1} \pi_{n1}  & \gamma_{n2} \pi_{n2} & \cdots & -\sum_{j, j \neq n}
\gamma_{nj} \pi_{nj}  \end{BMAT} \right]
\end{equation}
($\gamma_{ij}=0$ when $j \notin{\mathcal N}_{i}^{-}$), where $\pi_{ij}$ is unconditional probability that the node $j$ broadcasts and node $i$ updates its parameters as a consequence ($\pi_{ij}= \pi_{j} p_{ij}$, where $\pi_{j}$ is the unconditional probability for node $j$ to broadcast).

According to (\ref{algg}) and Lemma~\ref{lm1}, we shall consider $B(k)=P^{-\zeta} A \Gamma(k)$ and $\bar{B}=E \{B(k) \}= P^{-\zeta} A \bar{\Gamma}$ ($P^{-\zeta}= \bar{N}^{\zeta}
{\rm diag} \{ p_{1}^{-\zeta}, \ldots, p_{n}^{-\zeta} \})$.

\begin{lemma} \label{lm2} \cite{calieee}
Matrix $\bar{B}$ has one eigenvalue at the origin, and the remaining ones in the left half plane. Let $T= \left[
\begin{BMAT}{c.c}{c} \mathbf{1} & T_{n \times (n-1)} \end{BMAT} \right]$, where $T_{n \times (n-1)}$ is such that $ {\rm span} \{T_{n \times (n-1) }\}= {\rm span} \{\bar{B}\}$, while $\mathbf{1}=[1\ldots 1]^T$. Then,
\begin{equation} \label{t}
T^{-1} \bar{B} T= \left[ \begin{BMAT}{c.c}{c.c} 0 & 0_{1 \times (n-1)}   \\ 0_{(n-1) \times 1} &
\bar{B}^{*}
\end{BMAT} \right],
\end{equation}
where $\bar{B}^{*}$ is Hurwitz.
\end{lemma}
Consequently, there exists $R^{g}>0$ satisfying
\begin{equation} \label{lyap}
 R^{g} \bar{B}^{*} +
\bar{B}^{*T}R^{g}=-Q^{g},
\end{equation}
for any given $Q^{g} > 0$.
It also follows from the derivation of \eqref{t} that
$
T^{-1} B(k) T=\left[ \begin{BMAT}{c.c}{c.c} 0 & B_{1}(k) \\ 0_{(n-1) \times 1} &
B_{2}(k)
\end{BMAT} \right],$ with $E\{B_{1}(k)\}=0$ and $E \{B_{2}(k) \}=\bar{B}^{*}$.
\par
Properties of the random variable $\Delta t(k)$ are important for further analysis of (\ref{algg}).
\begin{lemma} \label{lm3}
$E\{ \Delta t(k) \}=\frac{1}{\mu_{j}} \frac{l-m}{p_{ij}}$, ${\rm var} \{\Delta t(k)\}=\frac{1}{\mu_{j}^{2}}\frac{l-m}{p_{ij}}$, where $l-m=L$ for \emph{AlgDrift.a}, $l-m=\lfloor(1-\nu)l\rfloor$ for \emph{AlgDrift.b} and $l-m=l$ for \emph{AlgDrift.c}; for large $l$, we have $l \sim \pi_{ij} k $.
\end{lemma}

\subsection{Convergence of Drift Correction Algorithm}

After coming back to (\ref{algg}), we first insert $\varepsilon^a(k)$ from (\ref{Delta}). Then, we introduce
 $\tilde{g}(k)=T ^{-1} \hat{g}(k)$ and decompose $\tilde{g}(k)$ as  $\tilde{g}(k)=[\tilde{g}(k)^{[1] }  \; \vdots \; \tilde{g}(k)^{[2] T}]^{T}$, where $\tilde{g}(k)^{[1]}=\tilde{g}_{1}(k)$
 and $\tilde{g}(k)^{[2]}=[\tilde{g}_{2}(k) \ldots \tilde{g}_{n}(k)]^{T}$. After neglecting the higher order terms from (\ref{Delta}), we obtain
\begin{align}
  \tilde{g}(k+1)^{[1]}  = &\tilde{g}(k)^{[1]}+\frac{1}{k^{\zeta}} F_{1}(k)\Delta t(k) \tilde{g}(k)^{[2]}  +\frac{1}{k^{\zeta}} H_{1}(k)\tilde{g}(k)   \label{g1} \\
   \tilde{g}(k+1)^{[2]} = &\{I+ \frac{1}{k^{\zeta}}[\bar{B}^{*} +F_{2}(k)] \Delta t(k)\} \tilde{g}(k)^{[2]}  + \frac{1}{k^{\zeta}} H_{2}(k) \tilde{g}(k),  \label{g2}
\end{align}
where matrices $F_{1}(k)$ and $F_{2}(k)$ are defined by
\[T^{-1} [  B(k) - \bar{B}^{*}] T= \left[ \begin{BMAT}{c.c}{c.c}  0 & F_{1}(k) \\ 0_{(n-1) \times 1} & F_{2}(k) \end{BMAT} \right],\] while $H_{1}(k)$ and $H_{2}(k)$ are defined by $T^{-1}  P^{-c} N_{g}(k) T=
 \left[ \begin{BMAT}{c}{c.c}  H_{1}(k) \\  H_{2}(k) \end{BMAT} \right].$

We now have the following convergence result for the drift correction algorithm.

\begin{theorem} \label{th1}  Let assumptions (A1)--(A3) be satisfied. Then,  $\tilde{g}(k)^{[1] }$ from \eqref{g1} converges to a random variable $\chi^{*}$ with bounded second moment,
and $\tilde{g}(k)^{[2] }$ from \eqref{g2} to zero in the mean square sense and w.p.1; in other words, $\hat{g}(k)$ generated by
(\ref{algg}) converges for all three choices of $m$ to $ \hat{g}_{\infty}=\chi^{*} \mathbf{1 }$ in the mean square sense and w.p.1. \end{theorem}

The rate of convergence of the drift estimation scheme is of utmost importance not only for the convergence of local clocks to a common virtual clock, but also for the convergence of the offset estimation algorithm. Asymptotic rate of convergence to consensus of the algorithm (\ref{algg}) will be studied through the behavior of $\tilde{g}(k)^{[2]}$ in (\ref{g2}), using the methodology of \cite[Chapter~3]{hfchen}.
\par
\begin{theorem} \label{th2} Let (A1)--(A3) hold. Then, $z(k)=k^{\zeta d} \tilde{g}(k)^{[2]}$, where $d > 0$ and $\tilde{g}(k)^{[2]}$ is defined by
(\ref{g2}), converges to zero in the mean square sense and w.p.1, when $\zeta' < 1$ for:
\newline
-~~ $\zeta d < \zeta'-\frac{1}{2}$ (\emph{AlgDrift.a}),
\newline
-~~ $\zeta d < \frac{1}{2}+\zeta'$ (\emph{AlgDrift.b}) and
\newline
-~~ $\zeta d < \zeta'$ (\emph{AlgDrift.c}),
\newline
and when $\zeta'=1$ for:
\newline
-~~ $d < \min ( \frac{1}{2}$, $2q r)$ (\emph{AlgDrift.a}),
 \newline
-~~ $ d < \min ( \frac{3}{4}$, $q r)$ (\emph{AlgDrift.b}) and
 \newline
-~~ $d <\min ( \frac{1}{2}$, $q r)$ (\emph{AlgDrift.c}),
\newline
where $r=\frac{\lambda_{min}(Q^{g})}{\lambda_{max}(R^{g})}$,  $q=\frac{L}{\max_{i,j}(\mu_{j} p_{ij})}$ for \emph{AlgDrift.a}, $q=\frac{1-\nu}{\mu_{c}}$ for \emph{AlgDrift.b} and $q=\frac{1}{\mu_{c}}$ for \emph{AlgDrift.c}.
\end{theorem}

\begin{remark}  The given conditions are sufficient and, generally, conservative; the results hold asymptotically, for $k$ large enough. They indicate that the \emph{AlgDrift.b} gives the best results: in the case when $\zeta' < 1 $, the important result $\zeta d > 1 $ is achieved, enabling convergence to a common virtual clock (see Corollary~\ref{cor1} below).  This is a consequence of the variable left end of the intervals $[m,l]$, which introduces a white noise term in the recursion (\ref{gi}) (see the Theorem proof); at the same time, unbounded increase of interval length~ $l-m$ ensures an effectively increasing signal-to-noise ratio and appropriate averaging. \emph{AlgDrift.c} with fixed $m$ does not allow this effect. However, in practice, it is sufficient to choose $l-m=L$ large enough and to apply \emph{AlgDrift.a}, avoiding in such a way practical problems connected with the unbounded increase of memory inherent to \emph{AlgDrift.b}. It will be demonstrated in Section~\ref{sim} by simulation that, practically, the best results can be obtained by \emph{AlgDrift.a} for $L$ moderately high.

  Notice that the  convergence rate $\zeta d > 1$, important for achieving convergence to a global virtual clock, is not achievable by the CBTS algorithms discussed in \cite{scfi,tian1,tian2,tian3}. This is not a contradiction w.r.t. the result of Theorem~2, having in mind that the algorithms are structurally different.
\end{remark}

An important conclusion resulting from Theorems~\ref{th1} and \ref{th2} is that
\begin{equation} \label{g}
\hat{g}(k)=
\chi(k)  \mathbf{1} + \hat{g}(k)^{[2]}, \; {\rm w.p.1}
 \end{equation}
where $\chi(k)= \tilde{g}(k)^{[1]}$ and $\hat{g}(k)^{[2]}= T_{n \times (n-1)} \tilde{g}(k)^{[2]}$, with  $\chi(k)= \chi^{*}+o(1)$ and $\|\hat{g}(k)^{[2]}\|= o(\frac{1}{k^{\zeta d}})$. The last relation is fundamental for the convergence analysis of the offset correction estimation.

\subsection{Convergence of Offset Correction Algorithm}

We start the analysis by introducing the following expressions in (\ref{algf}) and (\ref{algc}):
\begin{eqnarray} \label{expr}
&\Gamma(k)=\bar{\Gamma}+ \tilde{\Gamma}(k), \; \; \Gamma_{d}(k)=\bar{\Gamma}_{d}+ \tilde{\Gamma}_{d}(k),& \nonumber \\ & \xi^{0}(k)=\bar{\xi}^{0}+ \tilde{\xi}^{0}(k), \;\; \eta^{0}(k)=\bar{\eta}^{0}+ \tilde{\eta}^{0}(k),& \\ & \bar{\delta}(k)=\bar{\delta^{0}}+ \tilde{\delta}(k), \;\; t^{0}(k)=\bar{t}^{0}+\tilde{t}^{0}(k), & \nonumber
\end{eqnarray}
 where $\bar{\Gamma}= E\{ \Gamma(k) \}$, $\bar{\Gamma}_{d} = E \{ \Gamma_{d}(k) \}$, $\bar{\xi}^{0} = E \{\xi^{0}(k) \}= \sum_{j=1}^{n} \xi(t^{j,i}_{0}) \pi_{j},$ $\bar{\xi}^{0}_{d} = {\rm diag} \,\bar{\xi}^{0}$,  $\bar{\eta}^{0} =  E \{\eta^{0}(k) \}= \sum_{j=1}^{n} \eta(\bar{t}^{j,i}_{0}) \pi_{j}$, \[\bar{\delta}^{0} =E \{\bar{\delta}(k)\}= \sum_{j=1}^{n} [\bar{\delta}^{1,j}_{0} \ldots \bar{\delta}^{n,j}_{0}]^{T} \pi_{j}\] and  $\bar{t}^{0}=E \{t^{0}(k) \}=\sum_{j=1}^{n} t^{j,i}_{0} \pi_{j}$. Therefore, $\{ \tilde{\Gamma}(k) \}$, $\{\tilde{\Gamma}_{d}(k)\}$,  $ \{\tilde{\xi}^{0}(k)\}$, $\{\tilde{\eta}^{0}(k)\}$, $\{\tilde{\delta}^{0}(k)\}$ and $\{ \tilde{t}^{0}(k)\}$ are zero mean i.i.d. random sequences (due to randomness in determining the transmitting node for a given $k$).

\begin{theorem} \label{th3} Let assumptions (A1)--(A3) be satisfied and let $\hat{g}(k)$ be generated by \emph{AlgDrift.a} with $\zeta' \in (\frac{3}{4},1)$, and by \emph{\emph{AlgDrift.b}} or \emph{AlgDrift.c} with $\zeta' < 1$. Then,  $\hat{f}(k)$, generated by \textit{AlgOffset.a} using (\ref{algf}), converges to $\hat{f}^{*}$ and $\hat{c}(k)$ from (\ref{algc}) converges to $\hat{c}^{*}$ in the mean square sense and w.p.1 for all $\zeta'' \in (\frac{1}{2},1]$ in the case of \emph{AlgDrift.b} and \emph{AlgDrift.c}, and for all $\zeta''  \in (\frac{1}{2},1]$, $\zeta'' > \frac{3}{2}-\zeta'$, in the case of \emph{AlgDrift.a}; $\hat{f}^{*}$ and $\hat{c}^{*}$ satisfy the equation
\begin{equation} \label{m10}
 [\bar{\Gamma} \vdots \bar{\Gamma}_{d}] \hat{h}^{*}=0,
 \end{equation}
where $\hat{h}^{*}=[(\hat{f}^{*}+\chi^{*}\bar{\xi}_{d}^{0} A^{-1} \mathbf{1})^{T} \vdots (\hat{c}^{*}-\chi^{*} A (\bar{\eta}^{0}+\bar{\delta}))^{T}]^{T}$.
\end{theorem}

\begin{remark}
Rate of convergence of $\hat{g}(k)$ to consensus plays an important role in the offset correction algorithm. It influences $\hat{f}(k)$ in (\ref{algf}) directly, through the term $\Delta \hat{g}(k+1)$, and indirectly, through the remaining terms depending on $\hat{g}(k)$. The standard offset estimation algorithms derived from the unmodified error function $\varphi^{b}_{i} (\bar{t}_{l}^{j,i})^{0}= \bar{\tau}_{j}(t_{l}^{j,i}) -\bar{\tau}_{i}(\bar{t}_{l}^{j,i})$ with its linearly increasing term (see Remark~\ref{rem_offset}) makes the vast majority of drift estimation algorithm inapplicable in the case of measurement noise. According to Theorem~\ref{th3}, all the proposed algorithms for drift estimation can be utilized under appropriate assumptions. Theorem~\ref{th3} holds for any $\hat{g}(k)$ providing sufficient convergence rate to consensus, according to (\ref{g}).
\end{remark}

\begin{theorem} \label{th4}
 Let the assumptions of Theorem~\ref{th3} hold. Then $\hat{f}(k)$ and $\hat{c}(k)$, generated by the algorithm \textit{AlgOffset.b} ((\ref{algf}), (\ref{algc}) with consensus iterations on $\hat{c}(k)$ using \eqref{consensus}), converge in the mean square sense and w.p.1 to
$\hat{f}^{*}$ and  $\hat{c}^{*}=\hat{c}^{con} \mathbf{1}$, respectively ($\hat{c}^{con}$ is a scalar), where $\hat{f}^{*}$ and $\hat{c}^{con}$ satisfy the equation $M_{1}^{con} \hat{h}^{con}=0,$
where
\begin{equation} \label{m1con}
M_{1}^{con}=
\left[ \begin{BMAT}{c.c}{c.c} \bar{\Gamma} & {\rm vec} \{ \bar{\Gamma}_{d} \}  \\  - \sum_{i=1}^{n} \bar{\phi}_{i} \bar{\Gamma}^{(i)}   & -\sum_{i=1}^{n} \bar{\phi}_{i} {\rm vec} \{\bar{\Gamma}_{d} \}_{i}   \end{BMAT} \right],
\end{equation}
$\hat{h}^{con}=[(\hat{f}^{*}+\chi^{*}\bar{\xi}_{d}^{0} A^{-1} \mathbf{1})^{T} \vdots \hat{c}^{con}-\sum_{i=1}^{n} \bar{\phi}_{i} \chi^{*} (A \bar{\eta}^{0}+A \bar{\delta})_{i}]^{T}$, with $\bar{\phi}=[\bar{\phi}_{1} \ldots \bar{\phi}_{n}]$, $\bar{\phi} \bar{C}=\bar{\phi}$ and $\bar{C}=E\{ C(k) \}$; $\bar{\Gamma}^{(i)}$ denotes $i$-th row of the matrix  $\bar{\Gamma}$, and ${\rm vec} \{\bar{\Gamma}_{d} \}_{i}$ $i$-th element of ${\rm vec} \{\bar{\Gamma}_{d} \}$.
\end{theorem}

\begin{remark} \label{cons} Theorems~\ref{th3} and \ref{th4} specify the convergence points for the proposed offset correction algorithms. They depend explicitly (in the definition of $ \hat{h}^{*}$) not only on delays and measurement noise properties, but also on the convergence point of the drift estimation algorithm. They also depend on the network properties and the a priori selected weights through the relations (\ref{m10}) and (\ref{m1con}). In general, the corrected offsets in both \textit{AlgOffset.a} and \textit{AlgOffset.b} do not converge to the same point for all the nodes. However, a comparison between (\ref{m10}) and (\ref{m1con}) indicates clearly that it can achieve lower dispersion of the components of $\hat{f}^{*}$ within $\hat{h}^{con}$ due to lower number of degrees of freedom, implied by the consensus scheme. Simulation results presented in Section~\ref{sim} confirm this statement. 
\end{remark}

\subsection{Special Cases}

 When communication delays and measurement noise can be neglected, the algorithm \textit{AlgOffset.b} ((\ref{algg}), (\ref{algf}), (\ref{algc}) with \eqref{consensus}) is able \emph{to achieve consensus} on both corrected drifts  $\hat{g}_{i}(k)$ and corrected offsets $\hat{f}_{i}(k)$. Namely, in this case we have
\begin{align}
 \bar{\Gamma} \hat{f}^{*}  + \bar{\Gamma}_{d} \mathbf{1 } \hat{c}^{con}&=0  \nonumber \\
\sum_{i=1}^{n} \bar{\phi}_{i} \{ \bar{\Gamma}^{(i)} \hat{f}^{*}   + (\bar{\Gamma}_{d} \mathbf{1})_{i} \hat{c}^{con} \}&=0
\end{align}
The equation $\bar{\Gamma} \hat{f}^{*}  = -\bar{\Gamma}_{d} \mathbf{1 } \hat{c}^{con}$ has a nontrivial solution for $\hat{f}^{*}$ only for $\hat{c}^{con}=0$, having in mind that $\bar{\Gamma}_{d} \mathbf{1 }$ does not belong to the column space of $\bar{\Gamma}$. Therefore, we have $\bar{\Gamma} \hat{f}^{*}=0$, wherefrom the result follows.
\par
However, according to Theorem~\ref{th3}, \emph{AlgOffset.a} does not guarantee convergence of $\hat{f}(k)$ to consensus, due to additional degrees of freedom in the solution of $[\bar{\Gamma} \vdots \bar{\Gamma}_{d}] \hat{h}^{*}=0$. Elimination of the recursion for $\hat{c}(k)$ leads to divergence of offset estimates.
\par
When the stochastic terms $\xi(\cdot)$ and $\eta(\cdot)$ are equal to zero, it is possible to achieve exponential convergence rate by adopting constant step size in \emph{AlgDrift.a}, \emph{AlgOffset.a} and \emph{AlgOffset.b}, and   $\varepsilon_{i}^a(k)=\varepsilon' \nu_{i}(k)^{-1}$ in \textit{AlgDrift.b} and \emph{AlgDrift.c}. However, the offset correction algorithm again does not provide consensus, in general.
\par
When, in addition, the delay is equal to zero, the algorithm can be further simplified.  Assuming that \emph{AlgDrift.a} is applied and that $\bar{\varphi}_{i}^{b} (\bar{t}^{j,i})= \bar{\varphi}_{i}^{b} (\bar{t}_{l}^{j,i})^{0}= \bar{\tau} _{j}(t_{l}^{j,i})
-\bar{\tau} _{i}(\bar{t}_{l}^{j,i})$  for offset estimation, we come up with a synchronization algorithm in which $b_{i}$ is estimated using (\ref{bi}), where $\hat{\varphi}_{i}^{b}(\bar{t}_{l}^{j,i})=\hat{\tau}_{j}(t_{l}^{j,i}) -\hat{\tau}_{i}(\bar{t}_{l}^{j,i})$. The resulting algorithm is able to achieve \emph{exponential convergence to consensus} for both corrected drifts and offsets. This result is obtained for the first time in \cite{tsync2012} for pseudo periodic communication sequences (basically, convergence properties of such an algorithm are equivalent to those from \cite{scfi}).
When the delays are constant ($\delta^{j,i}_{l}=\bar{\delta}^{j,i}$), the offset estimates diverge. Introduction of a recursion for $\hat{c}(k)$ leads to convergence in the sense of Theorem~\ref{th3}.

\subsection{Common Virtual Clock} \label{common_clock}

As pointed out, the general aim of clock synchronization is convergence of local corrected times to a common virtual time. In view of the above results, we have:

\begin{corollary} \label{cor1} Let (A1)--(A3) be satisfied, with $\zeta'< 1$. Then, for \emph{AlgDrift.b} and either \emph{AlgOffset.a} or \emph{AlgOffset.b}, $\sup_{i,j} $ $\Delta \hat{\tau}_{i,j}(k)=\hat{\tau}_{i}(k)-\hat{\tau}_{j}(k)$ is bounded in the mean square sense and w.p.1. \end{corollary}

\begin{remark} Since
\begin{equation} \label{virtcl}
 \Delta \hat{\tau}_{i,j}(k)=[\hat{g}_{i}(k)-\hat{g}_{j}(k)] t(k)+ \hat{f}_{i}(k)-\hat{f}_{j}(k),
 \end{equation}
according to Theorem~\ref{th2}, we have that for $\zeta' < 1$ the first term at the right-hand side tends to zero only for \emph{AlgDrift.b}; for \emph{AlgDrift.a} and \emph{AlgDrift.c} convergence of $\Delta \hat{\tau}_{i,j}(k)$ is not achievable. As mentioned above, a practically realizable solution is to choose \emph{AlgDrift.a} with $L$ large enough (see simulation results). Notice also that all the estimates $\hat{f}_i(k)$ and the differences $\hat{f}_{i}(k)-\hat{f}_{j}(k)$ remain bounded by virtue of the adoption of the special error function in (\ref{jb}); otherwise, they diverge.
\end{remark}

\begin{remark}
As already stated, the overall approach in \cite{ggl1} suffers from the problem that $[\hat{g}_{i}(k)-\hat{g}_{j}(k)]$ does not tend to zero, allowing unbounded increase of the first term in (\ref{virtcl}). The control-based approach from \cite{ycs} attempts to reduce $\Delta \hat{\tau}_{i,j}(k)$ directly by a careful choice if PI regulator parameters; however, the given analysis does not provide an insight into overall network stability.
\end{remark}

\subsection{Tuning Network Weights and the Flooding Scheme} \label{flooding}
Coefficient $\gamma_{ij}$ in (\ref{ai}) and (\ref{bi}) is the weight of the update at node $i$, occurring as a consequence of a tick at node $j$, $i,j=1, \ldots, n$. If one wishes to express high confidence in the precision of a given clock, there are two basic implementations: 1) to increase either all the elements $\gamma_{ji}$, $i=1, \ldots, n$,  or the Poisson rate $\mu_{j}$ for a given $j$; 2) to decrease the weights $\gamma_{ij}$, $j=1, \ldots, n$,  for a given $i$. The first way clearly gives more weight to the sender.The second way is related to the receiver, implying lower increments of the local parameter changes at node $i$, and, therefore, lower influence of the rest of the network to the corresponding local parameter estimates. In the limit, node $i$ does not update its parameters ($\gamma_{ij}=0$, $j=1, \ldots, n$), and becomes a \emph{reference node} with fixed corrected parameters. The whole algorithm becomes in such a way an algorithm of \emph{flooding} type \cite{mksl,wuchse,suak}.
 \par
\begin{corollary} \label{cor2} Let the assumptions of Theorem~\ref{th1} be satisfied. Let node $\lambda$ be a \emph{center node} in $\mathcal{G}$, with the corrected drift $\hat{g}_{\lambda}^{*}$. Then, after setting ${\mathcal N}_{\lambda}^{-} = \emptyset$ (or $\gamma_{\lambda j} =0$, $j=1, \ldots, n$), algorithm (\ref{ai}) provides convergence of all the corrected drifts $\hat{g}_{i}(k)$, $i=1, \ldots, n$, $i \neq \lambda$, to $\hat{g}_{\lambda}$ in the mean square sense and w.p.1.
\end{corollary}

\section{Simulations} \label{sim}
Numerous simulation experiments have been undertaken in order to get a practical insight into the proposed distributed time synchronization algorithm. Different networks have been simulated with variable number of nodes.  The assumed network topology corresponds to a modification of Geometric Random Graphs \cite{gupta2000}. The nodes represent randomly spatially distributed agents within a square area. Initially, the nodes are assumed to be connected if their Euclidean distance is less than a predefined number: this results in an undirected graph. The obtained graph is modified in such a way as to transform a certain percentage (roughly 10 percent) of the original two-way communications into one-way communications. A program is developed for final optimization, which ensures, on the basis of additional modifications, that assumption (A1) is satisfied.  Parameters $\alpha_{i}$ and $\beta_{i}$ are randomly chosen in the intervals $(0.96, 1.04)$ and $(-0.2,0.2)$, respectively. Average communication delays $\bar{\delta}^{j,i}$ have been chosen to be 0.1, while  $\{\eta(k)\}$ and $\{\xi(k)\}$ have been simulated as zero-mean Gaussian white noise sequences with specified standard deviation $\sigma$. It has been adopted that $\zeta'=\zeta''=0.99$ and that the communication dropouts occur according to the probability $p_{ij}=0.9$.

\begin{figure}
    \centering
    \begin{subfigure}{0.5\columnwidth}
        \centering
        \includegraphics[width=\columnwidth]{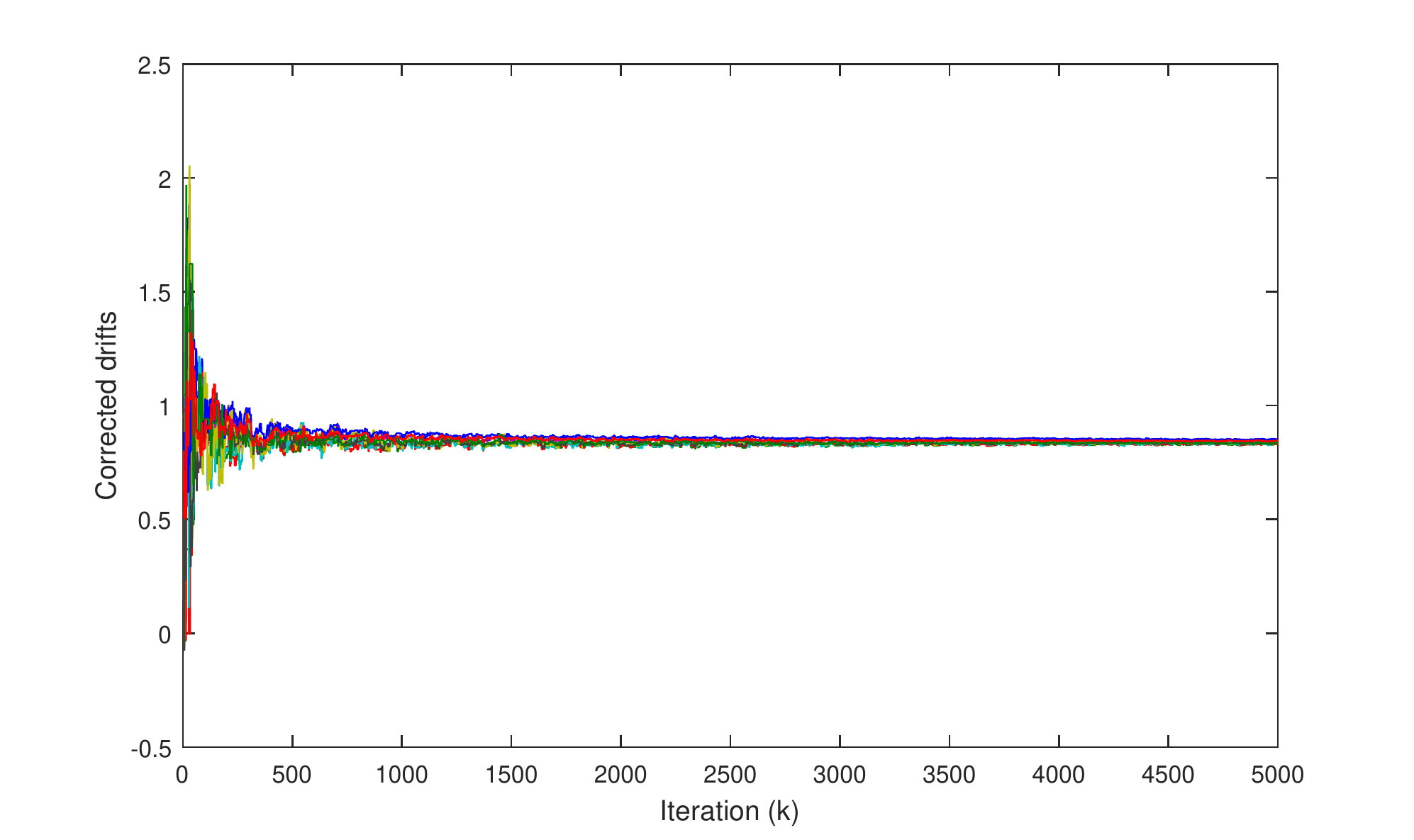}
        \caption{\small \textit{AlgDrift.a} with $L=1$}
    \end{subfigure}%
    \begin{subfigure}{0.5\columnwidth}
        \centering
        \includegraphics[width=\columnwidth]{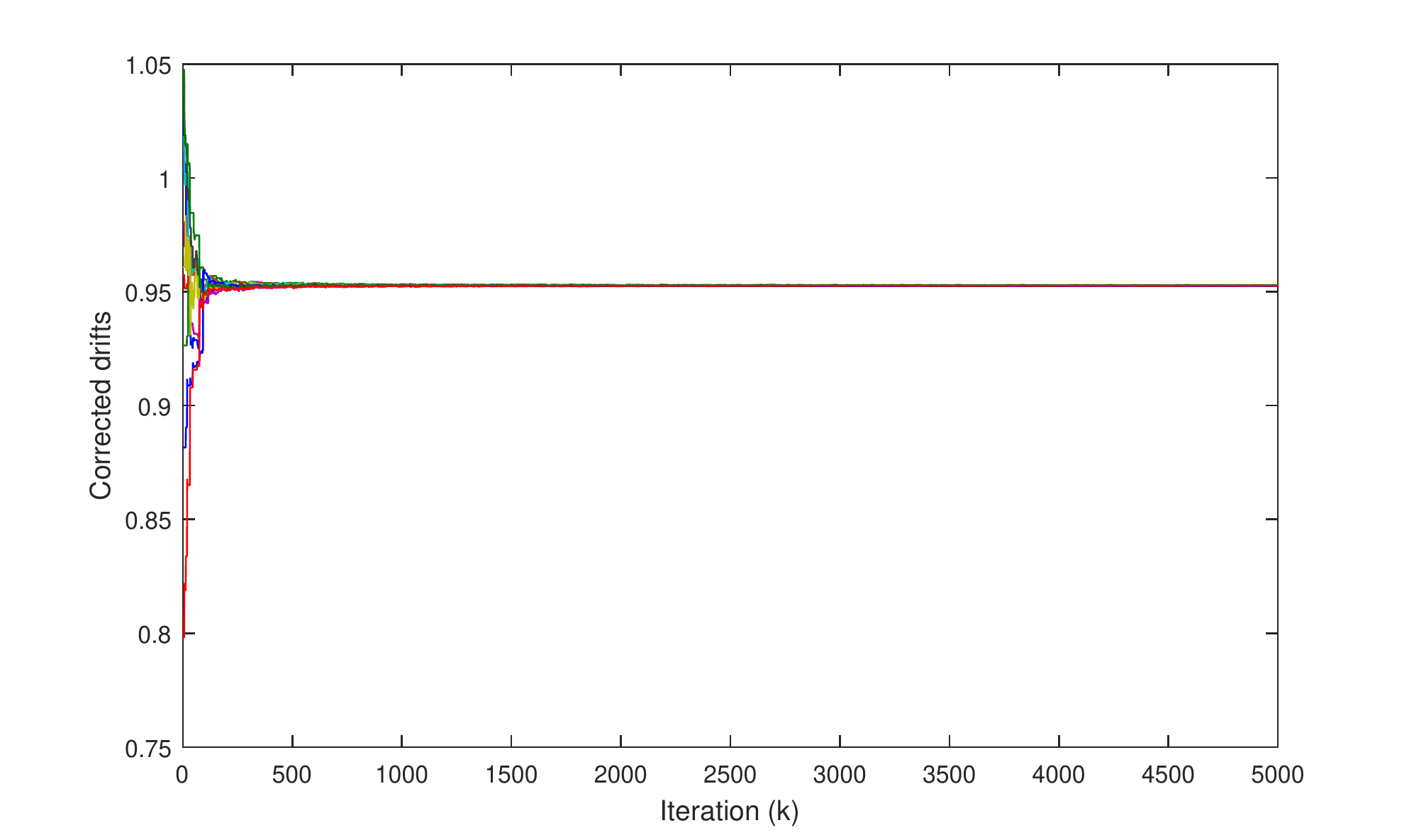}
        \caption{\small \textit{AlgDrift.a} with $L=100$}
    \end{subfigure}%

    \begin{subfigure}{0.5\columnwidth}
        \centering
        \includegraphics[width=\columnwidth]{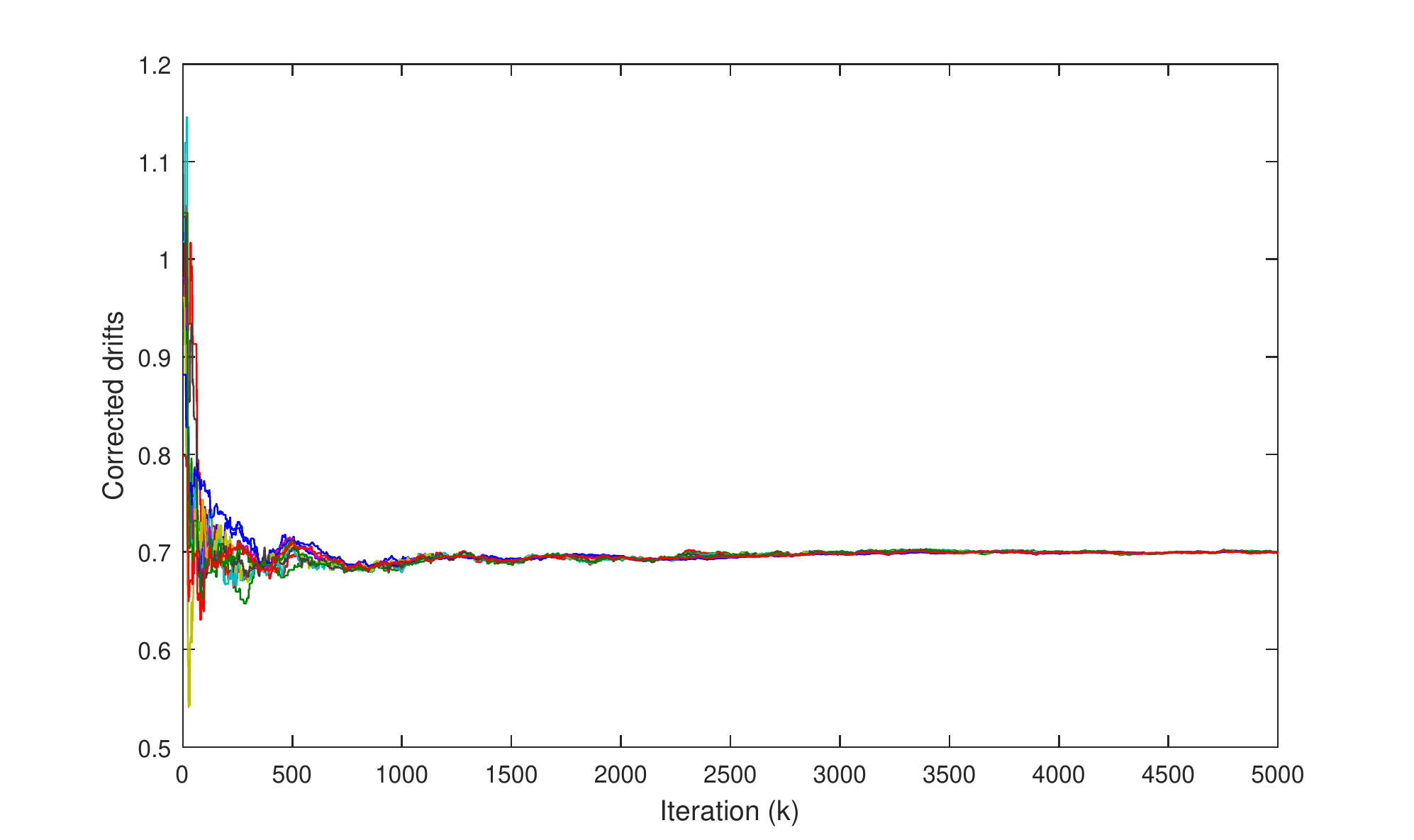}
        \caption{\small \textit{AlgDrift.b} with $\nu=1/2$}
    \end{subfigure}%
    \begin{subfigure}{0.5\columnwidth}
        \centering
        \includegraphics[width=\columnwidth]{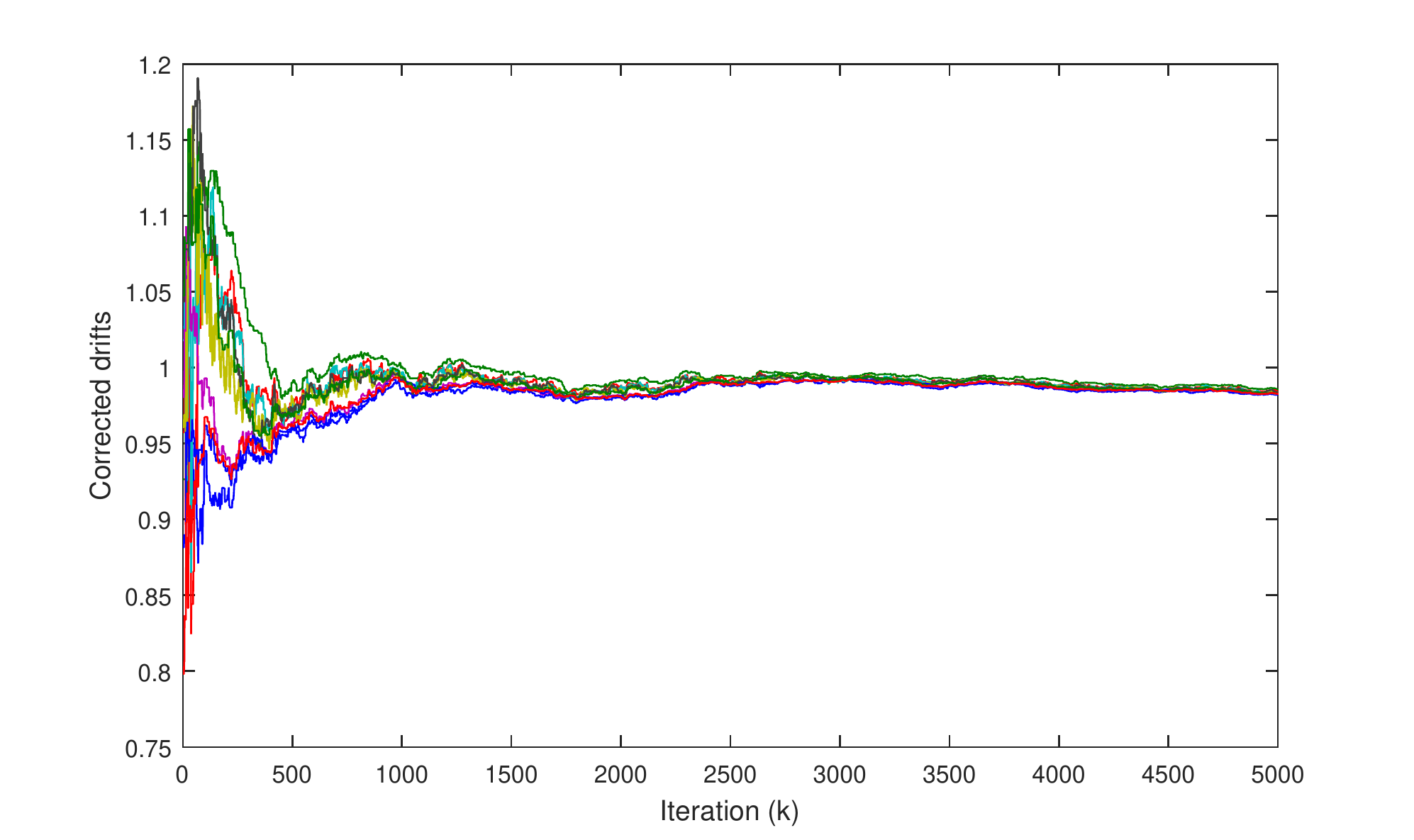}
        \caption{\small \textit{AlgDrift.c}}
    \end{subfigure}
    \caption{Corrected drifts: in spite of the theoretically proved asymptotic superiority of \textit{AlgDrift.b}, \textit{AlgDrift.a} with $L=100$ practically achives the best rate of convergence to consensus and noise immunity.} \label{drifts}
\end{figure}

Typical behavior of the corrected drifts generated by \emph{AlgDrift.a} ($L=1$ and $L=100$), \emph{AlgDrift.b} ($\nu=\frac{1}{2}$) and \emph{AlgDrift.c} ($l_{0}=0$) in the presence of stochastic delays and measurement noise with $\sigma = 0.05$  is presented in Fig.~\ref{drifts} for a network with ten nodes. Convergence to consensus can be clearly observed in all cases. Analogous schemes from the literature (\emph{e.g.}, \cite{tian1}) cannot achieve such a performance. The algorithm proposed in \cite{scfi} is very sensitive to noise and practically inapplicable under the given conditions, while the algorithm from \cite{tian2} achieves results similar to the ones obtained by \emph{AlgDrift.c}, but with typically lower convergence rate. It should be noticed that the best results are achieved by \emph{AlgDrift.a} with $L=100$; \emph{AlgDrift.b} is practically inferior on finite intervals, in spite of the asymptotic results from Theorem~\ref{th2}. This indicates that the best choice of drift estimation algorithm should be in practice connected to \emph{AlgDrift.a}, with a suitably selected $L$; it represents the best compromise between the signal to noise ratio and computational burden.

\begin{figure}
    \centering
    \begin{subfigure}{0.5\columnwidth}
        \centering
        \includegraphics[width=\columnwidth]{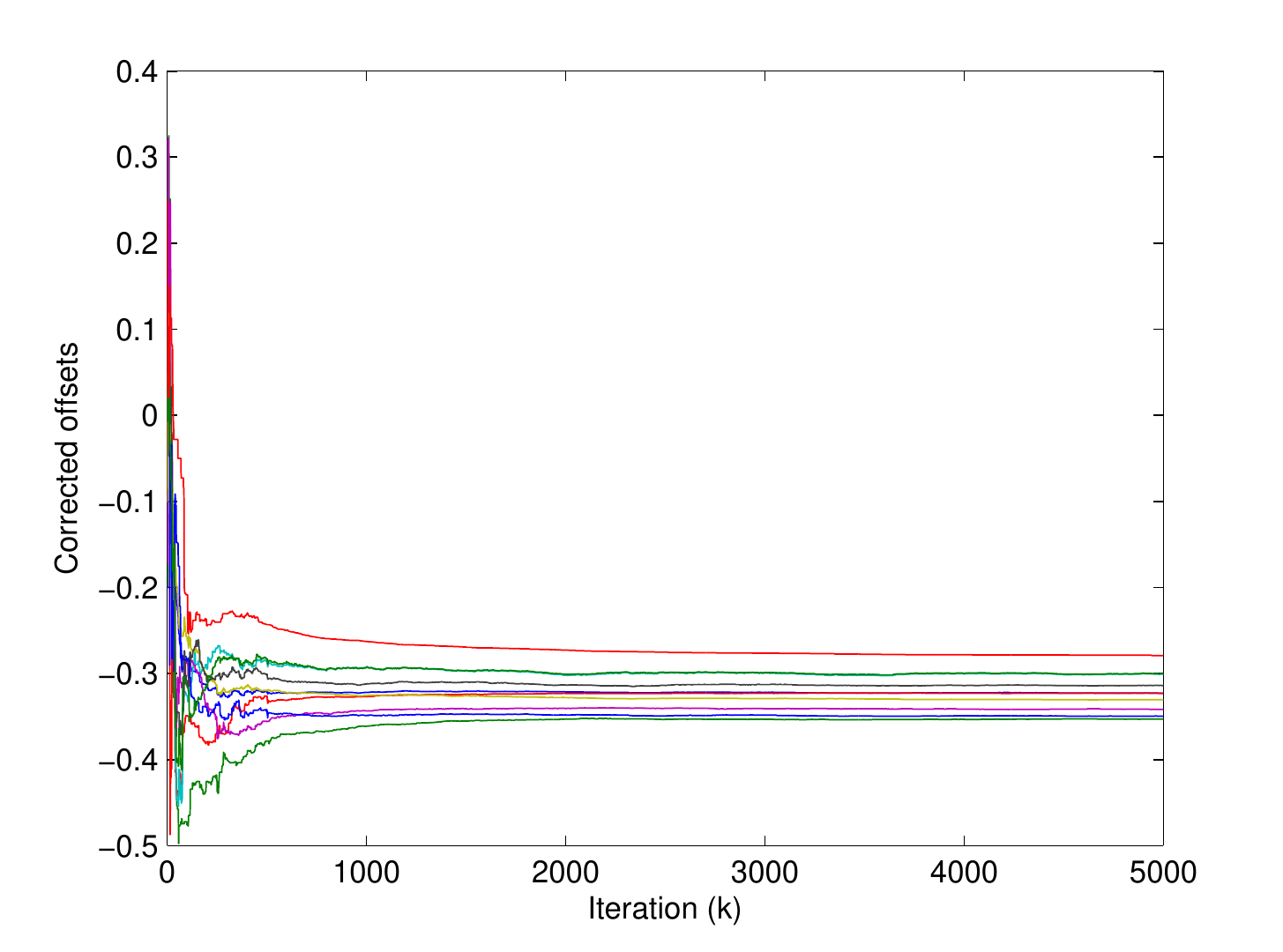}
        \caption{\small \textit{AlgOffset.a}}
    \end{subfigure}%
    ~
    \begin{subfigure}{0.5\columnwidth}
        \centering
        \includegraphics[width=\columnwidth]{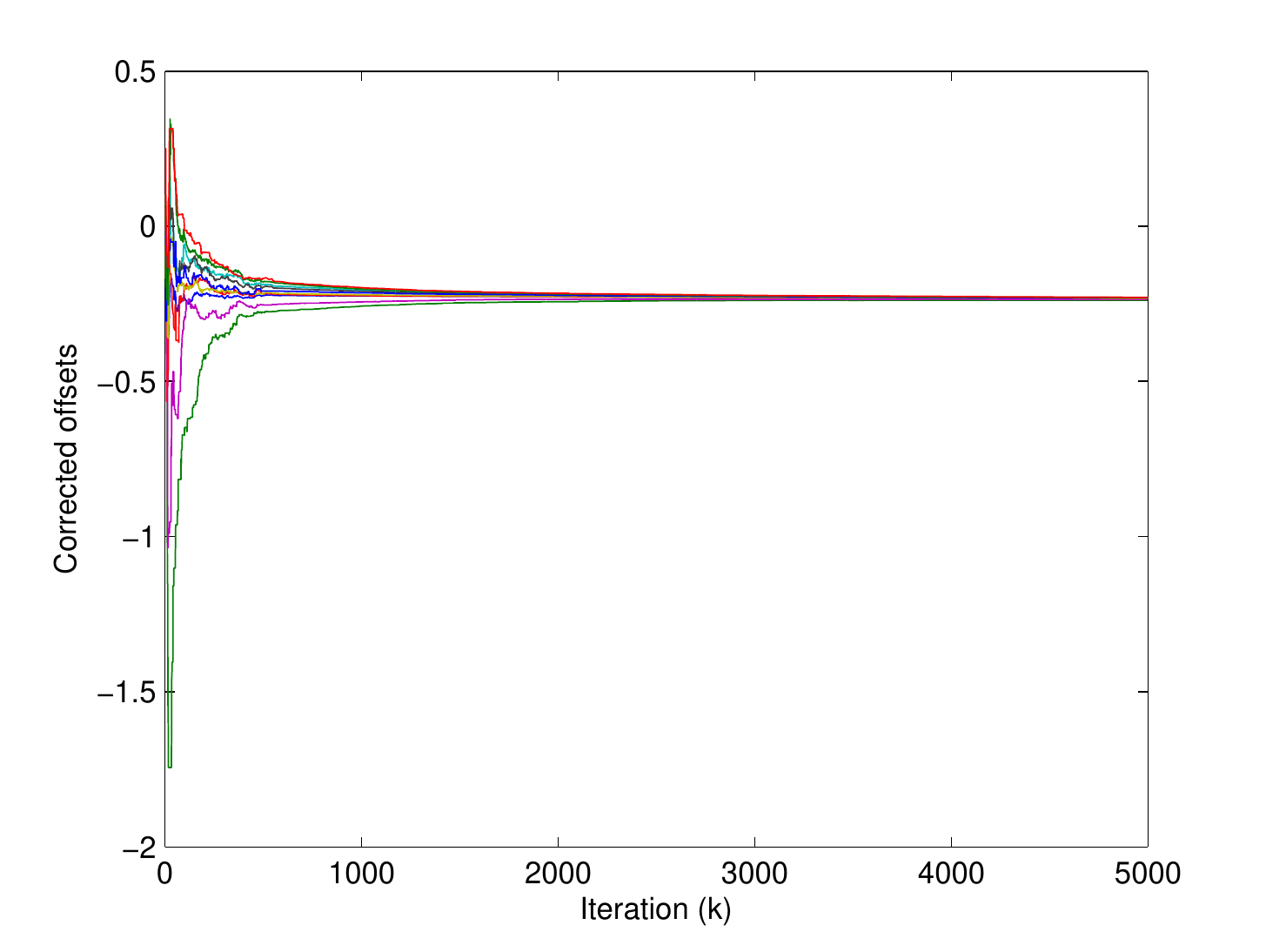}
        \caption{\small \textit{AlgOffset.b}}
    \end{subfigure}
    \\
    \begin{subfigure}{0.5\columnwidth}
        \centering
        \includegraphics[width=\columnwidth]{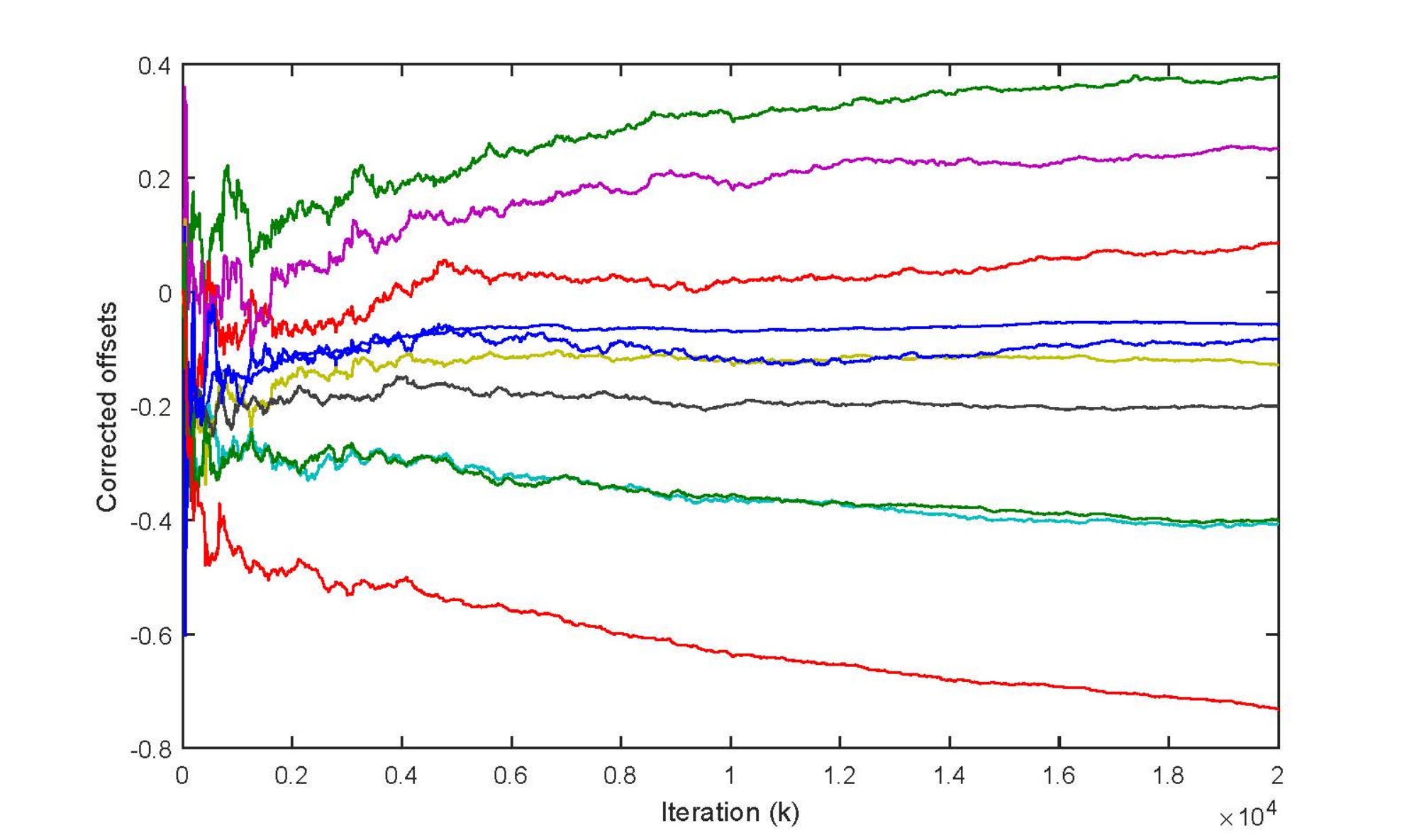}
        \caption{\small \textit{AlgOffset.a} with $T_{i}=T_j=0$}
    \end{subfigure}%
    ~
    \begin{subfigure}{0.5\columnwidth}
        \centering
        \includegraphics[width=\columnwidth]{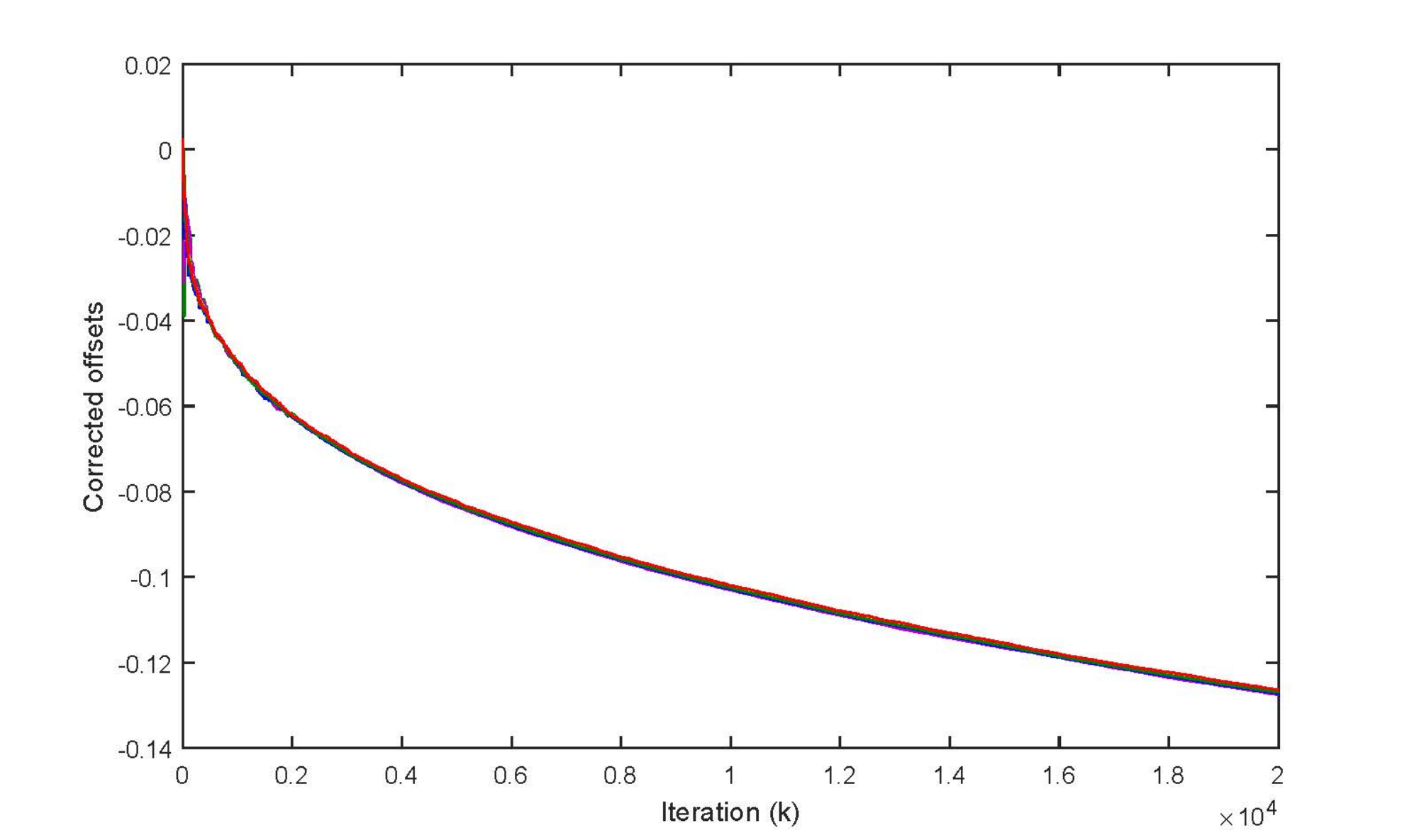}
        \caption{\small \textit{AlgOffset.a} with $c_{i}=0$}
    \end{subfigure}
    \caption{Corrected offsets: \textit{AlgOffset.b}, which includes consensus iterations on $\hat{c}(k)$, has lower dispersion of the converged estimates than \textit{AlgOffset.a}. Corrected offsets do not converge if we set $T_{i}=0$ (c) or $c_{i}=0$ (d), $i=1, \ldots, n$, illustrating the importance of the introduced modifications in the error function \eqref{jb}.} \label{offsets}
\end{figure}

Typical behavior of the proposed offset correction algorithms \emph{AlgOffset.a} and \emph{AlgOffset.b} is illustrated in Fig.~\ref{offsets} (a) and (b); \emph{AlgDrift.a} with $L=100$ has been used for drift correction. Convergence of all the components of the vector $\hat{f}(k)$ is evident in both cases.  The algorithm \textit{AlgOffset.b} provides a lower dispersion of the asymptotic values of the corrected offsets, as expected. Fig.~\ref{offsets} (c) and (d) illustrate the importance of introducing $T_{j}(\cdot)$ and $T_{i}(\cdot)$, given by \eqref{Tij}, and the delay correction parameter vector $c_{i}$ in the definition of  $\bar{\varphi}_{i}^{b} (\cdot)$ in (\ref{jb}), respectively. Fig.~\ref{offsets} (c) corresponds to $T_{j}(\cdot)=T_{i}(\cdot)=0$, and Fig.~\ref{offsets} (d) to $c_{i}=0$. It is evident that the offset estimates diverge in both cases. Introduction of $T_{j}(\cdot)$, $T_{i}(\cdot)$ and $c_{i}$ appears to be essential for obtaining convergence of the corrected offset estimates.

\begin{figure}
\begin{center}
\includegraphics[width=\columnwidth]{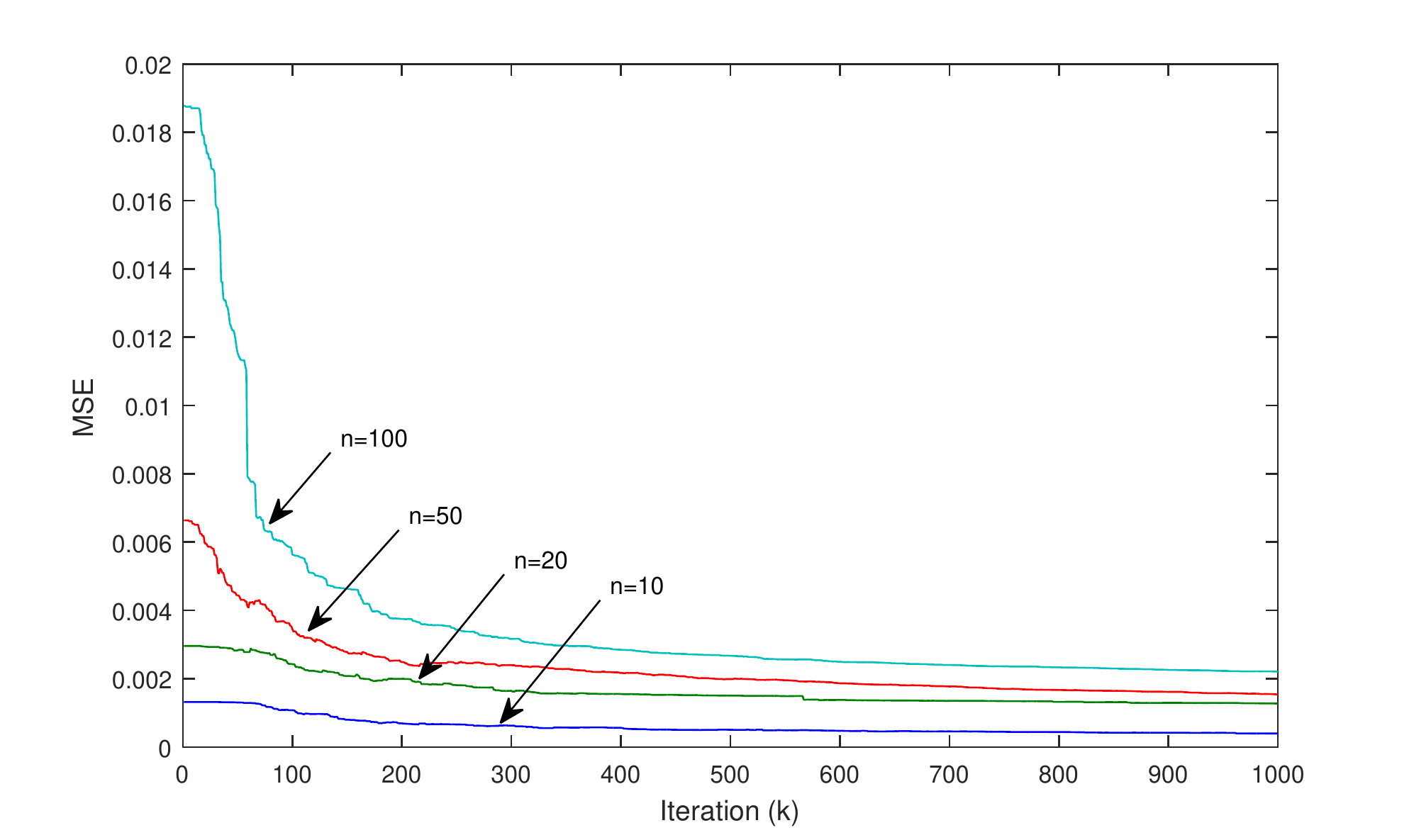}
\end{center}
\caption{Mean square disagreement for networks with 10, 20, 50 and 100 nodes.} \label{disagreement}
\end{figure}

In order to provide an insight into scalability of the proposed algorithm, in Fig.~\ref{disagreement} the mean square disagreement (the squared error between the local corrected drifts averaged over the number of nodes) is presented for networks generated at random by the above described procedure, and having 10, 20, 50 and 100 nodes. According to, \emph{e.g.},  \cite{bork}, it is possible to distinguish two regions in the figure. In the beginning of the first region, the disagreement between the nodes depends on the number of nodes \emph{almost linearly}. This is to be expected, having in mind that networks with approximately the same level of connectedness have been simulated. The convergence rate is fast and nearly exponential, dominantly influenced by the eigenvalue of the matrix $\bar{B}$ with the second smallest module. In the second part, when $k$ increases and $\nu_{i}(k)^{-\zeta'}$ tends to zero, all curves tend to zero, having in mind Theorem 1. The disagreement between the nodes increases with the number of nodes, but, typically, at a \emph{much slower rate}. As stated in Theorem 2, asymptotic convergence rate is characterized by $O(k^{-\zeta d})$, where the proportionality constant depends not only on the eigenvalues of the matrix $\bar{B}$, but also on the noise level. One should bear in mind that it is hard, in such an analysis, to separate the influence of the number of nodes from the network connectedness. Anyhow, the above consideration clearly shows that the method is indeed characterized by high scalability.
\begin{figure}
\begin{center}
\includegraphics[width=\columnwidth]{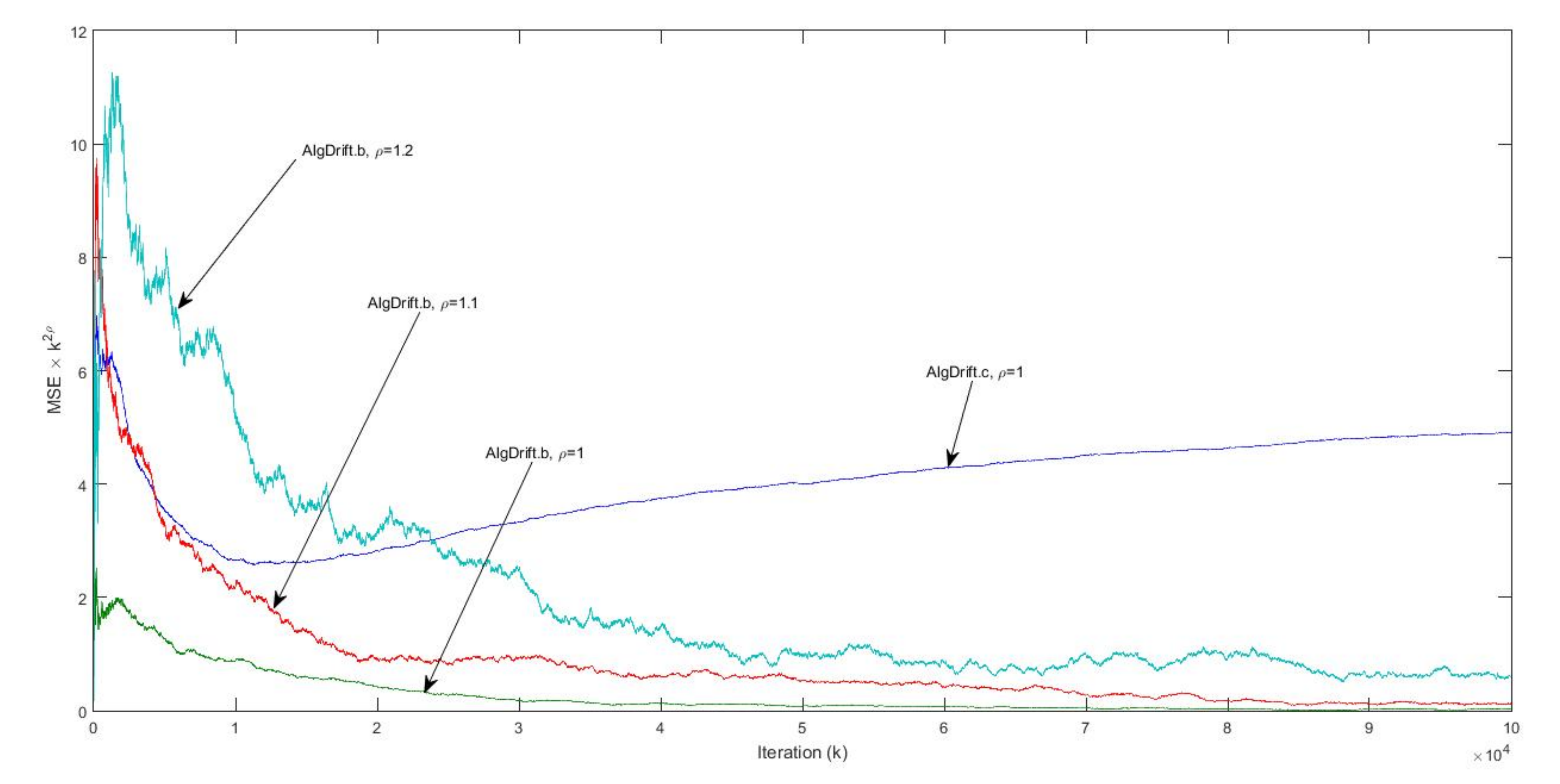}
\end{center}
\caption{Rate of convergence to a common virtual clock: the mean square disagreement multiplied by $k^{2\rho}$, for $\rho=1,1.1,1.2$ (for \emph{AlgDrift.b}) and $\rho=1$ (for \emph{AlgDrift.c}).} \label{rates}
\end{figure}

Fig.~\ref{rates} illustrates the rate of convergence to a common virtual clock (see Subsection~\ref{common_clock}): it represents the mean square disagreement multiplied by $k^{2\rho}$ where the exponent $\rho$ has been chosen to be 1, 1.1 and 1.2 for \emph{AlgDrift.b}, and $\rho=1$ for \emph{AlgDrift.c}, as indicated in the figure. The curve corresponding to \emph{AlgDrift.c} does not show convergence to a common virtual clock.

\section{Conclusion}
In this paper, a new distributed asynchronous algorithm has been proposed for time synchronization in networks with random communication delays, measurement noise and communication dropouts. A new algorithm is proposed for drift correction parameter estimation, based on an error function derived from specially defined local time increments. It has been proved, using the stochastic approximation arguments, that this algorithm achieves asymptotic consensus of the corrected drifts in the mean square sense and w.p.1, under general conditions concerning network properties. It is important that the algorithm achieves convergence rate superior to all similar schemes, especially in view of convergence to a virtual global clock. For offset estimation a new algorithm has been proposed starting from local time differences and adding: 1) special terms that take care of the influence of increasing time due to drift estimates, and 2) compensation parameters that take care of communication delays. It has been proved that the corrected offsets converge in the mean square sense and w.p.1 to finite random variables. An efficient algorithm for practical applications based on introducing consensus on compensation parameters has also been proposed. It has been also shown that the proposed algorithms can be used as flooding algorithms with one reference node. Simulation results provide illustrations of the presented theoretical results and confirm that the proposed algorithm represents an efficient tool for practice, outperforming all similar algorithms.

\appendix
\section{Proof of Lemma~\ref{lm1}}
According to (A2), $p_{i} > 0$, $i=1, \ldots, n$. Let $\tilde{k}$ correspond to the ticks of the global communication clock.  By Lemma~3 in \cite{ned}, for $\tilde{k}$ large enough, $\nu_{i}(\tilde{k})=\tilde{k} p_{i}+\chi_{i}(\tilde{k})$, where $|\chi_{i}(\tilde{k})| \leq \kappa \tilde{k}^{\frac{1}{2}+q'}$ w.p.1, $\kappa > 0$. Therefore,
\[ \nu_{i}(\tilde{k})^{-\zeta}= (\tilde{k} p_{i}+\xi_{i}(\tilde{k}))^{-\zeta}=(\tilde{k} p_{i})^{-\zeta}[1+ O(\frac{\chi_{i}(\tilde{k})}{\tilde{k}})] \]
Consequently, there exists $\tilde{\varepsilon}_{i} > 0$ such that  $|\frac{1}{\nu_{i}(\tilde{k})^{\zeta}} - \frac{1}{(\tilde{k} p_{i})^{\zeta}} | \leq \tilde{\varepsilon}'_{i} \frac{1}{\tilde{k}^{\zeta+\frac{1}{2}-q'}}$ w.p.1. The result of Lemma~\ref{lm1} follows  after taking into account that $\nu_{i}(k) = \nu_{i}(\tilde{k})$ for all iteration numbers $k$ between two consecutive updates at node $i$, and that $k \sim \tilde{k} \bar{N}$ for $\tilde{k}$ large enough. Formally, $E\{ N(j) | j \}$, the average number of updates for a broadcast from node $j$, can be obtained from the transmission probabilities $p_{ij}$, while $\bar{N}= \sum_{j=1}^{n} \pi_{j} E\{ N(j) | j \}$, where $\pi_{j}$ is the unconditional probability of node $j$ to broadcast. It is essential, however, that, asymptotically, $\nu_{i}(k)^{-\zeta}=O(k^{-\zeta})+O(k^{-\zeta-\frac{1}{2}+q'})$, where $q'>0$ is small enough.

\section{Proof of Lemma~\ref{lm3}} The result follows simply from the basic properties of the Poisson processes and the definition of the iteration number $k$.

\section{Proof of Theorem~\ref{th1}}

Introduce Lyapunov functions \[V^{g}(k)=E\{ (\tilde{g}(k)^{[1]})^{2} \}\] and \[W^{g}(k)= E\{
\tilde{g}(k)^{[2]T} R^{g} \tilde{g}(k)^{[2]} \},\] where $R^{g}>0$ satisfies \eqref{lyap} for a given $Q^{g} > 0$.
\par
In order to obtain an estimate of $V^{g}(k)$, we decompose $\tilde{g}(k+1)^{[1]}$ from (\ref{g1}) into the sum of zero input and zero state responses, defined by
\begin{equation}
\tilde{g}_{1}(k+1)^{[1]} = \Pi(k,1)^{[1]}\tilde{g}(1)^{[1]}
\end{equation}
 and
 \begin{align} \label{g21}
\tilde{g}_{2}(k+1)^{[1]}=\sum_{\sigma=1}^{k} \frac{1}{\sigma^{\zeta}} \Pi(k,\sigma+1)^{[1]}[F_{1}(\sigma) \Delta t(\sigma) +  H_{1}(\sigma)^{[2]}] \tilde{g}(\sigma)^{[2]},
\end{align}
respectively,
 where
$\Pi(k,l)^{[1]}=\prod_{\sigma=l}^{k} (1+ \frac{1}{\sigma^{\zeta}}H_{1}(\sigma)^{[1]})$,  $\Pi(k,k+1)^{[1]}=1$, and $H_{1}(k)^{[1]}$ follows from the decomposition $H_{1}(k)= [H_{1}(k)^{[1]} \vdots H_{1}(k)^{[2]}] $. Therefore, $V^{g}(k) \leq 2 V^{g}_{1}(k)+2 V^{g}_{2}(k)$, where $V^{g}_{1}(k)=E\{  (\tilde{g}_{1}(k)^{[1]})^{2} \}$ and $V^{g}_{2}(k)=E\{  (\tilde{g}_{2}(k)^{[1]})^{2} \}$.
\par
\par
Introduce $\sum_{i=1}^{n} |\mathcal{N}_{i}^{-}|$ infinite subsequences  $\{\kappa^{ij}(v)\} $ of the set of nonnegative integers $\mathcal{I}^{+}$, $i=1, \ldots, n$, $j \in \mathcal{N}_{i}^{-}$, $v=0, 1, 2, \ldots$, in which  $\kappa^{ij}(v)$ for a given $v$ defines an instant $k$ corresponding to an update at node $i$ realized as a consequence of a tick of node $j$ ($\kappa^{ij}(v_{1})$ $< \kappa^{ij}(v_{2})$ for $v_{1} < v_{2}$ and $\cup_{i,j} \{\kappa^{ij}(v)\} = \mathcal{I}^{+}$). 
Define $\Pi(k,1)_{s}^{[1]}=$ $ \prod_{\sigma \in  \{\kappa^{ij}(v)\}, \sigma \leq k } (1+ \frac{1}{\sigma^{\zeta}}H_{1}(\sigma)^{[1]})$, $s=1, \ldots, \sum_{i} |\mathcal{N}_{i}^{-} |$, $i=1, \ldots, n$, $j \in \mathcal{N}_{i}^{-}$, so that $\prod_{s} \Pi(k,1)_{s}^{[1]}= \Pi(k,1)^{[1]}$. According to the definition of $N^{g}(k)$, for \emph{AlgDrift.a} and \emph{AlgDrift.b}, the zero mean random sequences $\{ H_{1}(\sigma)^{[1]} \}$,  $\sigma \in \{\kappa^{ij}(v)\}$, have the property that $\{ H_{1}(\sigma)^{[1]} \}_{\sigma = \{\kappa^{ij}(v)\}}$ is correlated only with $\{ H_{1}(\sigma)^{[1]} \}_{\sigma = \{\kappa^{ij}(v-1)\}}$ and $\{ H_{1}(\sigma)^{[1]} \}_{\sigma = \{\kappa^{ij}(v+1)\}}$. Therefore, it follows that  $ E\{ (\Pi(k,1)_{s}^{[1]})^{2} \} < \infty$, since $\Pi(k,1)_{s}^{[1]}$ are mutually independent. For \emph{AlgDrift.c}, we have that $H_{1}(\sigma)^{[1]}=$ $ \tilde{H}_{1}(\sigma)^{[1]} - \tilde{H}_{1}(\sigma_0)^{[1]}$, where $\tilde{H}_{1}(\sigma)^{[1]}$ is zero mean i.i.d., and $\tilde{H}_{1}(\sigma_0)^{[1]}$ a finite w.p.1 random variable ($\sigma_{0}=\kappa^{i,j}(0)$). Therefore, we have
\begin{equation}
E \{ (1-\frac{1}{\sigma^{1+\zeta'}}H_{1}(\sigma))^{2}|\mathcal{F}_{\sigma_{0}} \} \leq 1-c_{1}\frac{1}{\sigma^{1+\zeta'}}+c_{2} \frac{1}{\sigma^{2(1+\zeta')}},
\end{equation}
where $\mathcal{F}_{\sigma_{0}}$ is the minimal sigma algebra generated by the measurements up to $\sigma_{0}$. It follows that $ E\{ (\Pi(k,1)_{s}^{[1]})^{2} \} < \infty$. Therefore, we obtain that $\sup_{k} V^{g}_{1}(k) < \infty$ for all three algorithms.
\par
Estimation of $V^{g}_{2}(k)$ for \emph{AlgDrift.a} and \emph{AlgDrift.b} starts from decomposing the sum at the right hand side of (\ref{g21}) into $\sum_{i=1}^{n} |\mathcal{N}_{i}^{-}|$ sums with indices $\sigma$ belonging to  $\{\kappa^{ij}(v)\} $, $\sigma \leq k$.
All these sums contain weighted zero mean random variables $F_{1}(\sigma) \Delta t(\sigma)+H_{1}(\sigma)^{[2]}$; their correlation with $\tilde{g}(\sigma)$ is w.p.1 of the order of magnitude of $\frac{1}{\sigma^{2c}}$, so that it can be neglected for $k$ large enough w.r.t. the corresponding terms in the expression for $V_{2}^{g}(k)$. It is important to notice that $\sum_{\sigma} E \{ \frac{1}{\sigma^{2 \zeta}} F_{1}(\sigma)^{2} \Delta t(\sigma)^{2} \} \leq \infty$ for \emph{AlgDrift.a} and \emph{AlgDrift.b}, by virtue of Lemma~\ref{lm3}. Noticing also that $\sup_{k} \Pi(k,1)^{[1]} < \infty$, it follows, after straightforward technicalities, that
 \begin{equation} \label{V1}
 V^{g}_{2}(k+1) \leq C_{1} \sum_{\sigma=1}^{k} \frac{1}{\sigma^{1+q''}} W^{g}(\sigma),
 \end{equation}
where $C_{1} > 0$ and $q'' > 0$. For \emph{AlgDrift.c}, the sum at the right hand side of (\ref{g21}) contains the terms $H_{1}(\sigma)^{[2]}= \tilde{H}_{1}(\sigma)^{[2]}- \tilde{H}_{1}(\sigma_0)^{[2]}$, $\sigma \in \kappa^{ij}(v)$.  Having in mind that $\{ \tilde{H}_{1}(\sigma)^{[2]} \}$ is zero mean  and $\tilde{H}_{1}(\sigma_0)^{[2]}$ is bounded w.p.1, for \emph{AlgDrift.c} $\sum_{\sigma} \frac{1}{\sigma^{2 \zeta}} E\{ \Delta t(k)^{2} \} < \infty$ by Lemma~\ref{lm3} and $\sum_\sigma \frac{1}{\sigma^{\zeta}} < \infty$, so that we obtain (\ref{V1}).
 \par
  Consequently,
 \begin{equation} \label{vgfin}
 V^{g}(k+1) \leq C_{2}[1+ \max_{1 \leq \sigma \leq k} W^{g}(\sigma)],
 \end{equation}
  where $C_{2} > 0$, having in mind that $\sum_{k=1}^{\infty} \frac{1}{\sigma^{1+q''}} < \infty$.
 \par
Estimation of $W^{g}(k)$ is based on considering the recursion (\ref{algg}) as a set of recursions on the sets $\{\kappa^{ij}(v)\} $, $i=1, \ldots, n$, $j \in \mathcal{N}_{i}^{-}$. We rewrite (\ref{g2}) for  $\{\sigma \in \kappa^{ij}(v)\} $ in the following way
\begin{equation}
\tilde{g}(\sigma+1)^{[2]} = \Pi(\sigma,\sigma)^{[2]}  \tilde{g}(\sigma)^{[2]}  + \frac{1}{\sigma^{\zeta}} H_{2}(\sigma)^{[1]} \tilde{g}(\sigma)^{[1]},
\end{equation}
where $\Pi(\sigma,\sigma)=I+ \frac{1}{\sigma^{\zeta}}[(\bar{B}^{*} +F_{2}(\sigma)) \Delta t(\sigma)+ H_{2}(\sigma)^{[2]}]$, while $H_{2}(\sigma)^{[1]}$ and $H_{2}(\sigma)^{[2]}$ follow from the decomposition $H_{2}(\sigma)= [H_{2}(\sigma)^{[1]} \vdots H_{2}(\sigma)^{[2]}] $.
\par
We start the analysis by observing that for any $n$-vector $x$ and any $\sigma$ large enough
\begin{equation} \label{xx}
x^{T}E \{ \Pi(\sigma, \sigma)^{[2]T} R^{g} \Pi(\sigma, \sigma)^{[2]} \}x \leq[1-\frac{2}{\sigma^{\zeta'}} q \frac{\lambda_{min}(Q^{g})}{\lambda_{max}(R^{g})} +O(\frac{1}{\sigma^{2\zeta'}})]x^{T} R^{g} x,
\end{equation}
where $0 < \lambda_{min}(Q^{g}),  \lambda_{max}(R^{g}) < \infty$ and $q=\frac{L}{\max_{i,j}(\mu_{j} p_{ij})}$ for \emph{AlgDrift.a}, $q=\frac{1-\nu}{\mu_{c}}$ for \emph{AlgDrift.b} and $q=\frac{1}{\mu_{c}}$ for \emph{AlgDrift.c}. As $q > 0$ (Lemma~\ref{lm3}), after standard technicalities based on the classical results on stochastic approximation \cite{hfchen,ky}, it follows that \[\prod_{\sigma \in \{\kappa^{ij}(v)\}} \| \Pi(\sigma, \sigma) \| \to_{\sigma \to \infty} 0,\] $i=1, \ldots, n$, $j \in \mathcal{N}_{i}^{-}$, in the mean  square sense and w.p.1, for \emph{AlgDrift.a}, \emph{AlgDrift.b} and \emph{AlgDrift.c}.
Moreover, as  $\{ H_{2}(\sigma)^{[1]} \}$ has the properties analogous to those of $\{ H_{1}(\sigma)^{[1]} \}$, it is possible to show, after technicalities similar to those utilized in the case of the analysis of $V^{g}(k)$, that  for $k$ large enough
\begin{equation} \label{wg}
W^{g}(k+1) \leq [1-c_{1}\frac{1}{k^{\zeta'}}] W^{g}(k)+C_{3}\frac{1}{k^{\zeta^{*}}} V^{g}(k),
\end{equation}
where $0 < c_{1}, C_{3} < \infty$, and
\newline
-~~ $\zeta^{*}=2 \zeta'$ for \emph{AlgDrift.a},
\newline
-~~ $\zeta^{*}=2(1+\zeta')$ for \emph{AlgDrift.b}, and
\newline
-~~ $\zeta^{*}=1+\zeta'$ for \emph{AlgDrift.c}.
\newline
Having in mind that $\sum_{k=1}^{\infty} k^{-\zeta^{*}} < \infty$ in all three cases, the methodology of \cite{huma3,huma5} can be applied, leading to the conclusion that
$\sup_{k} V^{g}(k) < \infty$.  Further, this gives rise to the conclusion that $\tilde{g}(k)^{[1]}$ tends to
a  random variable $\chi^{*}$ ($E\{ \chi^{*2} \} < \infty$)  and that $\tilde{g}(k)^{[2]}$ tends to zero in the mean square sense and
w.p.1. Consequently
$
\hat{g}_{\infty}= T \left[ \begin{BMAT}{c}{c.c}
\lim_{k \to \infty} \tilde{g}(k)^{[1]} \\ 0  \end{BMAT} \right]= \chi^{*} \mathbf{1},
$
which proves the theorem.

\section{Proof of Theorem~\ref{th2}}
After introducing the expression for $z(k)$ into (\ref{g2}),
we use the approximation $(1+\frac{1}{k})^{\zeta d} \approx 1+\zeta d \frac{1}{k}$ and obtain, after neglecting the higher order terms, that for $k$ large enough
\begin{align} \label{z2}
z(k+1)= z(k)+\{\frac{1}{k^{\zeta}} [\bar{B}^{*} +F_{2}(k)] \Delta t(k)  + \zeta d \frac{1}{k}I \}  z(k) + \frac{1}{k^{\zeta(1-d)}} H_{2}(k) \tilde{g}(k). 
\end{align}
Applying the methodology of the proof of Theorem~\ref{th1} to (\ref{z2}), we observe that for $\zeta' < 1$ the term proportional to $\frac{1}{k}$ can be neglected for $k$ large enough with respect to the term proportional to $\frac{1}{k^{\zeta'}}$.  We conclude from Theorem 1 that $\lim_{k \to \infty} z(k)=0$ in the mean square sense and w.p.1, provided, according to (\ref{wg}): a) $2 \zeta' (1-d) > 1$ for \emph{AlgDrift.a}, b) $2(1 +\zeta') (1-d) > 1$ for \emph{AlgDrift.b} and c) $(1+\zeta')(1-d) >\zeta'$ for \emph{AlgDrift.c}, wherefrom the first part of the result directly follows. Notice that different conditions result from different definitions of $\zeta$ and the properties of the corresponding sequence $\{ H_{2}(k) \}$. Inequality for \emph{AlgDrift.c} is more restrictive than the one for \emph{AlgDrift.b}, as a consequence of the fact that $\{ H_{2}(k) \}$ contains a term depending on the initial time $t_{l}^{0}$, which is fixed and nonzero for almost all realizations of the sequence $\hat{g}(k)$. Notice that it is possible to obtain a somewhat less restrictive condition for \emph{AlgDrift.c} using the inequality  $\zeta(1-d) > \zeta' \Rightarrow \zeta d < 1$, which follows from the analysis of the w.p.1 convergence using \cite{hfchen}, Chapter 3, Lemma 3.1.1, Theorem 3.1.1.
  \par
  For $\zeta'=1$, the terms proportional to $\frac{1}{k}$ and $\frac{1}{k^{\zeta'}}$ are of the same order of magnitude; as a result, the convergence conditions for (\ref{z2}) depend on the properties of the matrix $\bar{B}^{*}$. Hence the result follows.

\section{Proof of Theorem~\ref{th3}}

Let $\hat{h}(k)=[(\hat{f}(k)+\chi(k) \bar{\xi}_{d}^{0} A^{-1} \mathbf{1})^{T} \vdots (\hat{c}(k)- \chi(k)A (\bar{\eta}^{0}+\bar{\delta}))^{T}]^{T}$. As in the proof of Theorem~\ref{th1}, we obtain from (\ref{algf}), (\ref{algc}) and (\ref{expr}), after applying Lemma~\ref{lm1}, that
\begin{align} \label{h}
\hat{h}(k+1)=&\hat{h}(k)+\frac{1}{k^{\zeta''}} P_{d}^{-\zeta''} [ M_{1}(k)(\hat{h}(k) + u_{1}(k) \nonumber  \\ &+ u_{2}(k))+  M_{2}(k) \hat{G}(k)]- \frac{1}{k^{\zeta}}M_{3}(k)\hat{g}(k),
\end{align}
where \[u_{1}(k)= o(\frac{1}{k^{\zeta d}})[(A^{-1}\bar{\xi}^{0})^{T} \vdots (A(\bar{\eta}^{0}+\bar{\delta}))^{T}]^{T}, \]
 \[u_{2}(k)=[( \hat{g}_{d}(k)A^{-1} \tilde{\xi}^{0}(k))^{T} \vdots (\hat{g}_{d}(k)A (\tilde{\eta}^{0}(k) + \tilde{\delta}(k)))^{T} ]^{T}, \] $ M_{1}(k)= \bar{M}_{1}+
\tilde{M}_{1}(k),$ with
\[\bar{M}_{1}=\left[ \begin{BMAT}{c.c}{c.c} \bar{\Gamma} &   \bar{\Gamma}_{d}  \\  -\bar{\Gamma} & -\bar{\Gamma}_{d}    \end{BMAT} \right],
\tilde{M}_{1}(k)= \left[ \begin{BMAT}{c.c}{c.c} \tilde{\Gamma}(k) &  \tilde{\Gamma}_{d}(k)  \\ -\tilde{\Gamma}(k) &    -\tilde{\Gamma}_{d}(k)    \end{BMAT} \right], \]
$M_{2}(k)= \bar{M}_{2} + \tilde{M}_{2}(k)$,
$\bar{M}_{2}= {\rm diag} \{\bar{t}^{0} \bar{\Gamma}, \bar{t}^{0}  \bar{\Gamma} \},$  $\tilde{M}_{2}(k)= {\rm diag} \{\tilde{t}^{0}(k) \Gamma(k)+\bar{t}^{0} \bar{\Gamma}, \tilde{t}^{0}(k) \Gamma(k) +\bar{t}^{0} \bar{\Gamma} \}, $ $\hat{g}_{d}(k)= {\rm diag} \, \hat{g}(k)$,
$\hat{G}(k)=[\hat{g}(k)^{T} \vdots \hat{g}(k)^{T}]^{T}$ and $P_{d}^{-\zeta''}= {\rm diag} \{ P^{-\zeta''}, P^{-\zeta''} \}$; the last term in (\ref{h}) follows from  the term $\Delta \hat{g}(k+1)=\epsilon^{a}(k) [A \Gamma(k) \Delta t(k)+ N_{g}(k)] \hat{g}(k)$ in (\ref{algf}) and Lemma~\ref{lm1}, so that $M_{3}(k)=\left[ \begin{BMAT}{c}{c.c} P^{-\zeta} [A \Gamma(k) \Delta t(k)+ N_{g}(k)] \\ 0  \end{BMAT} \right]$.
 \par
From (\ref{h}) we realize that $P_{d}^{-\zeta''} \bar{M}_{1}$ has $n$ eigenvalues at the origin and $n$ eigenvalues in the left half plane. Therefore, there exists a nonsingular transformation $S$ such that
\begin{equation} \label{smin}
S^{-1} P_{d} ^{-\zeta''} \bar{M}_{1} S =\left[ \begin{BMAT}{c.c}{c.c} 0 & 0  \\  0 & \bar{M}^{*}   \end{BMAT} \right],
\end{equation}
 where $\bar{M}^{*}$ is Hurwitz (\cite{calieee}).
 Introduce $\tilde{h}(k)=S^{-1} \hat{h}(k)$, with $\tilde{h}(k)= [\tilde{h}(k)^{[1] T} $ $\vdots \tilde{h}(k)^{[2] T}]^{T}$, where ${\rm dim} \, \tilde{h}(k)^{[1] } = {\rm dim} \, \tilde{h}(k)^{[2] }=n$. Like in Theorem~\ref{th1}, we obtain from (\ref{h}) the following two recursions:
 \begin{align}
\tilde{h}(k+1)^{[1]}=&\tilde{h}(k)^{[1]}+\frac{1}{k^{\zeta''}} \{ \Psi(k)^{[1]} \tilde{h}(k)  \nonumber \\ &+ p(k)^{[1]}+q(k)^{[1]}+r(k)^{[1]} \}  \label{h1} \\
\tilde{h}(k+1)^{[2]}=&\tilde{h}(k)^{[2]}+\frac{1}{k^{\zeta''}} \{ \bar{M}^{*} \tilde{h}(k)^{[2]} + \Psi(k)^{[2]} \tilde{h}(k)  \nonumber \\ &+ p(k)^{[2]}+q(k)^{[2]}+ r(k)^{[2]}\},  \label{h2}
\end{align}
where \[S^{-1} P_{d}^{-\zeta''} \tilde{M}_{1}(k)S=\left[ \begin{BMAT}{c}{c.c}  \Psi(k)^{[1]} \\  \Psi(k)^{[2]} \end{BMAT} \right],\]
\[ S^{-1} P_{d}^{-\zeta''} [\tilde{M}_{1}(k) u_{1}(k)+M_{1}(k) u_{2}(k)+ \tilde{M}_{2}(k) \hat{g}(k)] = \left[ \begin{BMAT}{c}{c.c}  p(k)^{[1]} \\ p(k)^{[2]} \end{BMAT} \right],\]
\[S^{-1} P_{d}^{-\zeta''} [ \bar{M}_{1}(k) u_{1}(k)+ \bar{M}_{2} \hat{g}(k)]= \left[ \begin{BMAT}{c}{c.c}  q(k)^{[1]} \\  q(k)^{[2]} \end{BMAT} \right], \]
\[-S^{-1}   M_{3}(k) \hat{g}(k)= \left[ \begin{BMAT}{c}{c.c}  r(k)^{[1]} \\  r(k)^{[2]} \end{BMAT} \right]. \]
\par We introduce two main Lyapunov functions $V^{h}(k)=$ $ E\{ \| \tilde{h}(k)^{[1]} \|^{2} \}$ and $W^{h}(k)= $ $E\{ \tilde{h}(k)^{[2]T} R^{h} $ $ \tilde{h}(k)^{[2]} \}$, where $R^{h} > 0$ satisfies the Lyapunov equation $ R^{h} \bar{M}^{*} +
\bar{M}^{*T}R^{h}=-Q^{h}$, for any given $Q^{h} > 0$ (according to (\ref{smin})).
\par
At the first step, we set $q(k)^{[1]}=0$ and $q(k)^{[2]}=0$ and denote the corresponding solutions of (\ref{h1}) and (\ref{h2}) by $\tilde{h}_{1}(k)^{[1]}$ and $\tilde{h}_{1}(k)^{[2]}$, respectively. Then, we introduce \[V_{1}^{h}(k)=E\{ \| \tilde{h}_{1}(k)^{[1]} \|^{2} \}\] and \[W_{1}^{h}(k)= E\{ \tilde{h}_{1}(k)^{[2]T} R^{h} \tilde{h}_{1}(k)^{[2]} \}\]. It is straightforward to see that the results from \cite{huma3} can be directly applied to (\ref{h1}) and (\ref{h2}) (Theorem~11 and Lemma~12 therein), leading to the conclusion that $\sup_{k} V_{1}^{h}(k) < \infty$ and that $W_{1}^{h}(k)$ tends to zero when $k \to \infty$. It is essential for this conclusion that the sequences $\{\Psi(k)^{[1]}\}$, $ \{\Psi(k)^{[2]} \}$, $\{p(k)^{[1]}\}$ and $\{p(k)^{[2]}\}$ are uncorrelated, that $\sum_{k=1}^{\infty} \frac{1}{k^{2 \zeta''}} < \infty$, and that $\lambda_{min} (Q^{h}) > 0$.
\par

At the second step, consider the zero state responses $\tilde{h}_{2}(k)^{[1]}$ and $\tilde{h}_{2}(k)^{[2]}$ of (\ref{h1}) and (\ref{h2}) to the inputs $q(k)^{[1]}$ and $q(k)^{[2]}$, respectively. Let $V_{2}^{h}(k)=E\{ \| \tilde{h}_{2}(k)^{[1]} \|^{2} \}$ and $W_{2}^{h}(k)= E\{ \tilde{h}_{2}(k)^{[2]T} R^{h} \tilde{h}_{2}(k)^{[2]} \}$.
By (\ref{g}), we first conclude that
$
\bar{M}_{2} \hat{g}(k)=\bar{M}_{2}\hat{g}(k)^{[2]}
$ (having in mind Theorem~\ref{th2} and (\ref{g})).
From (\ref{h1}),  we obtain
\begin{equation} \label{h21}
\tilde{h}_{2}(k+1)^{[1]}=[I+\frac{1}{k^{\zeta''}} \Psi(k)_{1}^{[1]} ]\tilde{h}_{2}(k)^{[1]}+\frac{1}{k^{\zeta''}} q(k)^{[1]},
\end{equation}
where $\Psi(k)_{1}^{[1]}$ is an $(n \times n)$ submatrix of $\Psi(k)^{[1]}$. From (\ref{h21}) we have that $E \{\tilde{h}_{2}(k+1)^{[1]} \}= E \{\tilde{h}_{2}(k)^{[1]} \} + \frac{1}{k^{\zeta''}} q(k)^{[1]}; $ consequently,
\begin{align} V_{2}^{h}(k+1) \leq &(1+c' \frac{1}{k^{2\zeta''}}) V_{2}^{h}(k)+ (\frac{1}{k^{\zeta''}} q(k)^{[1]})^{2} \nonumber \\ &+ E \{\tilde{h}_{2}(k)^{[1]} \} \frac{1}{k^{\zeta''}} q(k)^{[1]},
\end{align}
($c' < \infty)$. Since $q(k)^{[1]}=o(\frac{1}{k^{\zeta d}})$ w.p.1, by Theorem~\ref{th2} we can derive that $\sup_{k} E \{\tilde{h}_{2}(k)^{[1]} \}^{2} < \infty$. Consequently,  $ \sup_{k} V_{2}^{h}(k) < \infty$ for all $\zeta'' > 1- \zeta d$, because of the requirement that $\sum_{k} \frac{1}{k^{\zeta''+\zeta d}} < \infty$. According to Theorem~\ref{th2}, when $\zeta' < 1$, this result holds for all $\zeta'' \in (\frac{1}{2},1]$ in the case of \emph{AlgDrift.b} and \emph{AlgDrift.c}, and for $\zeta''>\frac{3}{2}-\zeta'$ in the case of \emph{AlgDrift.a}. Analysis of $\tilde{h}_{2}(k)^{[2]}$ relies on the classical results from stochastic approximation \cite{hfchen}, wherefrom we obtain that $\lim_{k \to \infty} W_{2}^{h}(k)=0$.
\par
At the third step, consider the zero state responses $\tilde{h}_{3}(k)^{[1]}$ and $\tilde{h}_{3}(k)^{[2]}$ of (\ref{h1}) and (\ref{h2}) to the inputs $r(k)^{[1]}$ and $r(k)^{[2]}$, respectively. Let $V_{3}^{h}(k)=E\{ \| \tilde{h}_{3}(k)^{[1]} \|^{2} \}$ and $W_{3}^{h}(k)= E\{ \tilde{h}_{3}(k)^{[2]T} R^{h} \tilde{h}_{3}(k)^{[2]} \}$. Let $r_{1}(k)^{[1]}$ be the part of  $r(k)^{[1]}$ following from $\varepsilon^{a}(k) A \Gamma(k) \Delta t(k) \hat{g}(k)$, and $r_{2}(k)^{[1]}$ the part following from $\varepsilon^{a}(k) N_{g}(k) \hat{g}(k) $. Taking into account (\ref{g}), one concludes that $r_{1}(k)^{[1]} \sim o( \frac{1}{k^{\zeta'+\zeta d}})$ and that $r_{2}(k)^{[1]}$ is a zero mean i.i.d. term on any subsequence $\kappa^{ij}$, multiplied by $\frac{1}{k^{\zeta}}$. Therefore, $V_{3}^{h}(k) < \infty$, provided $\zeta'+ \zeta d > 1$; this is fulfilled in the case of \emph{AlgDrift.a} for $\zeta' > \frac{3}{4}$, and for any $\zeta' \in (\frac{1}{2},1]$ in the case of \emph{AlgDrift.b} and \emph{AlgDrift.c}. Reasoning similarly, we conclude that $\lim_{k \to \infty} W_{3}^{h}(k)=0$ under less restrictive conditions.

Therefore, we conclude  that $\sup_{k} V^{h}(k) < \infty$ and $\lim_{k \to \infty} W^{h}(k)=0$. Using the arguments exposed in \cite{huma3}, we further derive that $\tilde{h}(k)^{[1]}$ tends to a random $n$-vector $\tilde{h}^{[1]*}$, and that $\tilde{h}(k)^{[2]}$ tends to zero in the mean square sense and w.p.1, implying that $\hat{h}^{*}= \lim_{k \to \infty} \hat{h}(k)=S \tilde{h}^{*},$ where $\tilde{h}^{*}=[\tilde{h}^{[1]*T} \vdots 0_{1 \times n}^{T}]^{T}$.  The result of the theorem follows after taking into account that $\chi(k) \to  \chi^{*}$ w.p.1 and that \[\bar{M}_{1} \hat{h}^{*}= \bar{M}_{1} S \tilde{h}^{*}=P_{d}^{\zeta''}S\left[ \begin{BMAT}{c.c}{c.c} 0 & 0  \\  0 & \bar{M}^{*}   \end{BMAT} \right] \tilde{h}^{*}=0,\] according to (\ref{smin}) and the definition of $\hat{h}^{*}$.

\section{Proof of Theorem~\ref{th4}}
We  shall pay attention only to the possible convergence points: the rest can be derived by following methodologically the proof of Theorem~\ref{th3}. Namely, according to \cite{ky1}, we formulate the ODE characterizing the asymptotic behavior of the algorithm, and obtain that
\begin{align} \label{ode}
  \bar{\Gamma} \hat{f}^{*} - \chi^{*} [ \bar{\Gamma}_{d}  (\bar{\delta}+\bar{\eta}^{0})- \bar{\Gamma} \bar{\xi}^{0}] + \bar{\Gamma}_{d} \mathbf{1 } \hat{c}^{con}&=0 \nonumber \\
 \sum_{i=1}^{n} \bar{\phi}_{i} \{ \bar{\Gamma}^{(i)} \hat{f}^{*} - \chi^{*} [( \bar{\Gamma}_{d} [\bar{\delta}+\bar{\eta}^{0}])_{i} -(\bar{\Gamma} \bar{\xi}^{0})_{i}] + (\bar{\Gamma}_{d} \mathbf{1})_{i} \hat{c}^{con} \}&=0 ,
\end{align}
wherefrom the result directly follows.

\bibliography{formifaccalib1}

\end{document}